\documentclass [%
 reprint,nofootinbib,
superscriptaddress,
 amsmath,amssymb,
 aps,
]{revtex4-1}
\makeatletter
\renewcommand{\fnum@figure}{Fig. \thefigure}

\makeatother

\usepackage{soul}
\usepackage{xcolor}
\usepackage{graphicx}% Include figure files
\usepackage{dcolumn}% Align table columns on decimal point
\usepackage{bm}% bold math
\usepackage[T1]{fontenc}
\usepackage{subcaption}
\usepackage{mwe}
\usepackage{chngcntr}
\begin{document}

\preprint{APS/123-QED}

\title{High-Energy Phase Diagrams with Charge and Isospin Axes \\under Heavy-Ion Collision and Stellar Conditions}

\author{K. Aryal}
\affiliation{Department of Physics, Kent State University, Kent, OH 44243 USA}
\author{C. Constantinou}
\affiliation{Department of Physics, Kent State University, Kent, OH 44243 USA}
\author{R. L. S. Farias}
%\email{ricardo.farias@ufsm.br}
\affiliation{Departamento de F\'{\i}sica, Universidade Federal de Santa Maria,
97105-900 Santa Maria, RS, Brazil}
\author{V. Dexheimer}
%\email{vdexheim@kent.edu}
\affiliation{Department of Physics, Kent State University, Kent, OH 44243 USA}

\date{\today}% It is always \today, today,

\begin{abstract}
We investigate the phase transition from hadron to quark matter in the general case without the assumption of chemical equilibrium. The effects of net strangeness on charge and isospin fractions, chemical potentials, and temperature are studied in the context of the Chiral Mean Field (CMF) model that incorporates chiral symmetry restoration and deconfinement. The extent to which these quantities are probed during deconfinement in conditions expected to exist in protoneutron stars, binary neutron-star mergers, and heavy-ion collisions is analyzed via the construction of 3-dimensional phase diagrams.
\end{abstract}                       
                              
\maketitle

%\tableofcontents

\section{Introduction}

Recent works have discussed the possible similarities between the conditions in energetic astrophysical environments, such as protoneutron stars, core-collapse supernovae, and neutron-star binary mergers, to those present in heavy-ion collisions (HICs) \cite{Hanauske:2017oxo,Bastian:2018wfl}. These similarities are a consequence of (i) new low-energy heavy-ion experiments that are starting to produce matter at larger densities, together with more refined HIC calculations that allow for large chemical potential fluctuations at all energies \cite{Martinez:2019jbu} and (ii) high temperatures (compared to the Fermi temperature) achieved in some astrophysical phenomena that, unlike in the case of cold-catalyzed neutron stars, cannot be ignored. In particular, the temperature in protoneutron stars can be as high as $30-40$ MeV \cite{Burrows:1986me,Pons:1998mm} and, in mergers, it can exceed $50$ MeV or even reach $100$ MeV \cite{Galeazzi:2013mia,Perego:2019adq}. On the other hand, contemporary full general-relativity simulations \cite{Most:2019onn} indicate that neutron-star mergers cannot attain the large charge fractions of close to $Y_Q=0.4$ produced in HICs (eg. for Au-Au and Pb-Pb) and supernovae  \cite{Huedepohl:2009wh,Fischer:2009af}. This is a new feature as, before the advent of compact star mergers, all known hot astrophysical systems out of chemical equilibrium were newly formed and, therefore, still contained a significant amount of protons from the original heavy nuclei in the progenitors. The knowledge of such a large variety of conditions created the need to study hot and dense matter under a large range of charge fractions.

Phase diagrams for high energy matter (usually referred to as or Quantum Chromodynamics -~QCD~- phase diagrams) showing the position of the first-order deconfinement and chiral symmetry restoration phase transition are usually only depicted in two dimensions, temperature and baryon number density/chemical potential or temperature and isospin number density/chemical potential. The latter are interesting due to the fact that lattice QCD results are not afflicted by the sign problem at finite isospin chemical potential $\mu_I$~\cite{PhysRevD.66.034505,PhysRevD.66.014508,PhysRevD.98.094510,Brandt:2018wkp,Brandt:2016zdy,Brandt:2017zck,PhysRevD.97.054514}, as long as the baryon chemical potential $\mu_B$ remains zero. When $\mu_B\ne0$ or, equivalently, when there is a difference in the number of particles and anti-particles in the system, first-principle methods such as non-perturbative lattice QCD simulations cannot be performed due to the well-known fermion ``sign problem''~\cite{Karsch:2001cy,Muroya:2003qs,Bedaque:2017epw,Li_1998}. Although a considerable amount of theoretical work has been devoted to the subject \cite{PhysRevLett.86.592,Son:2000by,PhysRevD.99.096011,Carignano:2016lxe,PhysRevD.93.094502,Cohen:2015soa,PhysRevD.79.014021,PhysRevD.67.074034,PhysRevD.71.094001,PhysRevD.64.016003,Mannarelli:2019hgn,PhysRevD.64.016003,PhysRevD.95.105010,PhysRevD.98.054030,Khunjua:2018jmn,PhysRevD.88.056013,PhysRevD.82.056006,PhysRevD.79.034032,Andersen_2009,PhysRevD.75.096004,Ebert:2005wr,Ebert_2006,PhysRevD.74.036005,PhysRevD.71.116001,He:2005sp,PhysRevD.69.096004,Toublan:2003tt,Frank:2003ve,PhysRevD.75.094015,PhysRevC.89.064905,Andersen:2015eoa,PhysRevD.98.074016,Stiele:2013pma,PhysRevD.88.074006,Kamikado:2012bt,Parnachev:2007bc,Erdmenger:2007ja,Aharony:2007uu,Rebhan:2008ur,PhysRevD.89.076001,Nishihara:2014nsa,Lv:2018wfq,Adhikari:2019zaj,Adhikari:2019mlf,Adhikari:2019mdk,Avancini:2019ego,Lu:2019diy,Wu:2020knd}, the phase diagram of high-energy physics still remains poorly understood.

Another issue raised in the literature is the manner in which strangeness can affect phase diagrams. Ref.~\cite{Rennecke:2019dxt} has recently studied this for the particular case of isospin symmetric matter using functional renormalization theory, although this is not a new topic \cite{Lee:1986mn,Heinz:1987sj,Koch:1982ij,Lukacs:1986hu}. Lattice QCD calculations have also studied the effects of a non-zero strange chemical potential in, for example, the curvature of the chiral pseudo-critical line \cite{Bonati:2015bha,Borsanyi:2020fev}. In this work, we consider two scenarios. In one of them, there is no constraint on strangeness, assuming that chemical equilibrium with respect to the weak force has already been achieved, in which case there is no need to define a strange chemical potential. In the other case, it is assumed that there is not enough time for strangeness to be produced, in which case the strange chemical potential must be numerically determined in each phase to produce a zero net-strangeness fraction.

When the baryon chemical potential is finite, the usual practice in the literature has been to construct 2-dimensional phase diagrams (with temperature on the other axis) either in weak-chemical equilibrium, referring to fully-evolved neutron star matter with the charged chemical potential set to minus the electron chemical potential $\mu_Q = -\mu_e$, or in an isospin-symmetric configuration with $\mu_I = 0$, referring to matter created in relativistic HICs. In this work, we examine the behavior of the deconfinement coexistence region in 3-dimensional phase diagrams as a function of either the (hadronic and quark) charge fraction $Y_Q$ or the isospin fraction $Y_I$ (with temperature and baryon chemical potential/free energy completing our coordinate system). We do that numerically by varying the charged or isospin chemical potential, and in some cases also the strange chemical potential. Note that the charge fraction $Y_Q$ is the variable usually employed in equations of state for astrophysical applications, while the isospin fraction  $Y_I$ is more commonly used in HIC applications. The relation between the two quantities is trivial only when net strangeness is zero. Our calculations and discussion extend to phase diagrams in which the charge and isospin fractions are replaced by the corresponding chemical potentials, both at zero and non-zero net strangeness. As pointed out in Ref.~\cite{Monnai:2019hkn}, it is important to understand which plane of the high-energy phase diagram is being probed, eg. in the HIC Beam Energy Scan experiment, as the traditional critical point for isospin-symmetric matter without strangeness constraints may never be reached in the experiment.

\section{Formalism}
\subsection{The CMF model}
In order to construct our phase-diagrams, we make use of the Chiral Mean Field (CMF) description. It is based on a nonlinear realization of the SU(3) sigma model and constructed in such a way that chiral invariance is restored at large temperatures and/or densities. In its present version, it contains hadronic, as well as quark degrees of freedom \footnote{Note that an alternative version of the CMF model includes in addition the chiral partners of the baryons and gives the baryons a finite size \cite{Steinheimer:2011ea, Motornenko:2019arp}}, and its Lagrangian density is given by \cite{Dexheimer:2008ax, Dexheimer:2009hi}:
\begin{eqnarray}
\mathcal{L} = \mathcal{L}_{\rm{Kin}} + \mathcal{L}_{\rm{Int}} + \mathcal{L}_{\rm{Self}} + \mathcal{L}_{\rm{SB}} - U ,
\end{eqnarray}
where $\mathcal{L}_{\rm{Kin}}$ is the kinetic energy density of hadrons and quarks. The remaining terms are:
\begin{eqnarray}
\mathcal{L}_{\rm{Int}} &=& -\sum_i \bar{\psi_i} \big[\gamma_0 \big(g_{i \omega} \omega + g_{i \phi} \phi + g_{i \rho} \tau_3 \rho \big) + M_i^* \big] \psi_i , \nonumber\\
\mathcal{L}_{\rm{Self}} &=& \frac{1}{2} \big(m_\omega^2 \omega^2 + m_\rho^2 \rho^2 + m_\phi^2 \phi^2\big) \nonumber \\
&+& g_4 \left(\omega^4 + \frac{\phi^4}{4} + 3 \omega^2 \phi^2 + \frac{4 \omega^3 \phi}{\sqrt{2}} + \frac{2 \omega \phi^3}{\sqrt{2}}\right) \nonumber \\
&-& k_0 \big(\sigma^2 + \zeta^2 + \delta^2 \big) - k_1 \big(\sigma^2 + \zeta^2 + \delta^2 \big)^2 \nonumber \\
&-& k_2 \left(\frac{\sigma^4}{2}+\frac{\delta^4}{2} + 3 \sigma^2 \delta^2 + \zeta^4\right) - k_3 \big(\sigma^2 - \delta^2 \big) \zeta \nonumber 
\end{eqnarray}
\begin{eqnarray}
&-& k_4 \ \ln{\frac{ \big(\sigma^2 - \delta^2 \big) \zeta}{\sigma_0^2 \zeta_0}} , \\ \nonumber 
\mathcal{L}_{\rm{SB}} &=& -m_\pi^2 f_\pi \sigma - \left(\sqrt{2} m_k^ 2f_k - \frac{1}{\sqrt{2}} m_\pi^ 2 f_\pi\right) \zeta , \\ \nonumber \\
U &=& \big(a_o T^4 + a_1 \mu_B^4 + a_2 T^2 \mu_B^2 \big) \Phi^2 \nonumber \\
&+& a_3 T_o^4 \ \ln{\big(1 - 6 \Phi^2 + 8 \Phi^3 -3 \Phi^4 \big)} .
\label{upol}
\end{eqnarray}
Here, $\mathcal{L}_{\rm{Int}}$ represents the interactions between baryons (and quarks) mediated by the vector-isoscalar mesons $\omega$ and $\phi$ (strange quark-antiquark state), the vector-isovector $\rho$, the scalar-isoscalars $\sigma$ and $\zeta$ (strange quark-antiquark state), and the scalar-isovector $\delta$. $\mathcal{L}_{\rm{Self}}$ describes the self-interactions of the scalar and vector mesons. The chiral symmetry breaking term responsible for producing the masses of the pseudoscalar mesons is given by $\mathcal{L}_{\rm{SB}}$. U is the effective potential for the scalar field $\Phi$. It depends on the temperature and the baryon chemical potential in order to reproduce the standard view of the high-energy phase diagram concerning the shape of the deconfinement first-order phase transition coexistence line and its intersection with the zero-temperature axis. Its pure temperature contribution was fitted to reproduce the results of the Polyakov loop in the PNJL approach \cite{Ratti:2006ka, Roessner:2006xn} at zero baryon chemical potential (see details below when discussing quark couplings). The chemical potential and mixed terms were motivated by symmetry and simplicity. The former one also contains the correct scale in the asymptotic zero-temperature case. The index $i$ runs over the baryon octet  and the three light quarks. Leptons are not included in this calculation, since they are not present in HIC initial conditions and are not in chemical equilibrium with the rest of the system in the astrophysical scenarios we discuss.

The coupling constants of the hadronic part of the model are given in Ref.~\citep{Roark:2018uls}. They were fitted to reproduce the vacuum masses of baryons and mesons, nuclear saturation properties (density $\rho_0=0.15$ fm$^{-3}$, binding energy per nucleon $B/A=-16$ MeV, compressibility $K=300$ MeV), symmetry energy ($E_{\rm{sym}}=30$ MeV with slope $L=88$ MeV), and reasonable values for the hyperon potentials ($U_\Lambda=-28.00$ MeV, $U_\Sigma=5$ MeV, $U_\Xi=-18$ MeV). The predicted critical point for the nuclear liquid-gas phase transition of isospin symmetric matter lies at $T_c=16.4$ MeV, $\mu_{B,c}=910$ MeV. The vacuum expectation values of the scalar mesons are constrained by reproducing the pion and kaon decay constants.

As a result of their interactions with the mean field of mesons and the field $\Phi$, the baryons and the quarks acquire (Dirac) effective masses, which have the form:
\begin{eqnarray}
M_{B}^* &= g_{B \sigma} \sigma + g_{B \delta} \tau_3 \delta + g_{B \zeta} \zeta + M_{0_B} + g_{B \Phi} \Phi^2,  \nonumber \\
M_{q}^* &= g_{q \sigma} \sigma + g_{q \delta} \tau_3 \delta + g_{q \zeta} \zeta + M_{0_q} + g_{q \Phi}(1 - \Phi),
\end{eqnarray}
where the bare masses are $M_0=150$ MeV for nucleons, $354.91$ MeV for hyperons, $5$ MeV for up and down quarks, and $150$ MeV for strange quarks (see Ref.~\citep{Roark:2018uls} for the coupling constants in the quark sector). Notice that for vanishing values of $\Phi$, $M_{q}^*$ is large, which suppresses the quarks. Conversely, values of $\Phi$ close to $1$ suppress the hadrons. In this sense, $\Phi$ acts in our approach as an order parameter for deconfinement, as it only gives rise to a quark phase in the expected regime of large temperatures and/or densities.

The coupling constants of the quark sector are fitted to lattice data and to expectations from the phase diagram. The lattice data include (i) the location of the first-order phase transition and the pressure functional $P(T)$ at $\mu_B=0$ for pure gauge (the latter resulting from the PNJL model fitted to lattice) \cite{Ratti:2005jh, Roessner:2006xn} and (ii) the crossover pseudo-critical temperature and susceptibility $d\Phi/dT$ at vanishing chemical potential, together with the location of the ($T,\mu_B$) critical end-point for zero net-strangeness isospin-symmetric matter \cite{Fodor:2004nz}. The phase diagram expectations include a continuous first-order phase-transition line that starts at $T=167$ MeV temperature for zero-strangness isospin-symmetric matter and terminates on the zero-temperature axis at four times the saturation density of chemically-equilibrated and charge-neutral matter. See Ref.~\cite{Roark:2018uls,Dexheimer:2018dhb} for a detailed description of the effects of deconfined quarks inside neutron and protoneutron stars within the CMF model.

Note that the CMF description allows for the existence of a small admixture of quarks in the hadronic phase and a small admixture of hadrons in the quark phase at finite temperature. This feature becomes more prominent with increasing temperature and is required for the reproduction of the crossover transition known to take place at very large temperatures \cite{Aoki:2006we}. Despite this, quarks always have the dominant contribution in the quark phase (defined by a large $\Phi$), and hadrons in the hadronic phase (defined by a small $\Phi$). The first-order phase transition is characterized by a discontinuity in $\Phi$, which disappears at the critical point, beyond which (for higher temperatures) only one phase exists. Note that at zero temperature, the hadronic phase only contains hadrons ($\Phi=0$) and the quark phase only contains quarks ($\Phi\sim1$).

In this work, we choose to only show our phase diagrams up to $T = 160$ MeV, a little bit below the critical temperature $T_c = 167$ MeV predicted by the current parametrization of the model for zero net-strangeness isospin-symmetric matter. This is done for two different reasons. First, our critical point position was fitted and any modification to it would not affect the qualitative conclusions of our work. Second, we want to keep the discussion entirely general and the inclusion of a "special" feature, such as the critical point, would detract from our goals. In addition, the CompOSE \citep{CompOSE} repository contains equation of state tables that go up to $T = 160$ MeV, so all of our results could be reproduced as soon as our tables are uploaded to their website. So far, only the hadronic version of our tables are available online \citep{Dexheimer:2017nse}, but complete ones with quarks will be available soon.

\subsection{Useful Relations}  

We are interested in systems that are in equilibrium with respect to the strong and electromagnetic interactions, therefore, baryon number $B$ and electric charge $Q$ are conserved. In some of the cases we study, chemical equilibrium is not attained because weak interactions operate over much longer timescales (than the time scale of the system), introducing an extra condition of zero net strangeness $S$. The conserved quantities listed above correspond to our three independent chemical potentials $\mu_B$, $\mu_Q$, and $\mu_S$. The total chemical potential $\mu_i$ of each fermionic species $i$ can be expressed as a linear combination of these:
\begin{eqnarray}
\mu_i = Q_{B,i} \ \mu_B + Q_i \ \mu_Q +  Q_{S,i}\ \mu_S.
\label{equation:mui}\end{eqnarray}
This equation was derived in detail throughout Ref.~\cite{Hempel_2009} without any model assumption using Lagrange multipliers. The conventions we adopt for the values of the quantum numbers $Q$'s for the baryon octet and the three light quark species are given in Table I of Appendix A, followed by the resulting chemical potentials of the various species. Note that we consider the strangeness of particles to be positive in our notation, otherwise, all strangeness related quantities would have to have their signs reversed. For the purposes of our calculations, it is more convenient to work with fractions, the charge fraction being the amount of charged baryons and quarks over the total amount of baryons and quarks:
\begin{eqnarray}
Y_Q = \frac{Q}{B} = \frac{n_Q}{n_B^0}  = \frac{\sum_i Q_i \ n_i} {\sum_i Q_{B,i} \ n_i} ,
\label{6h}
\end{eqnarray}
where the $n's$ are number densities. Note that within the CMF model $n_B^0 = \sum_i Q_{B,i} n_i$ is not the same as the baryon number density $n_B$, as the latter comes from the derivative of the pressure with respect to the baryon chemical potential and, therefore, also contains a contribution from the potential $U$($\Phi$) when quarks are present (see Eq.~\eqref{upol} and discussion at the end of this section). For low temperatures, this contribution can be safely ignored on the hadronic-phase side of the phase-coexistence region, where $\Phi$ is approximately zero and, thus, $n_B^0 \simeq n_B$. Furthermore, we can insert the Gell-Mann-Nishijima relation \cite{Gell-Mann:1956iqa}: 
\begin{eqnarray}
Q_i = Q_{I,i} + {\frac{1}{2}}Q_{B,i}-{\frac{1}{2}}Q_{S,i},
\label{equation:qi}\end{eqnarray}
where $Q_{I,i}$ is the isospin of particle $i$, in the definition of charge density from Eq.~\eqref{6h} to obtain: 
\begin{eqnarray}
n_Q &= &\Sigma_i \left(Q_{I,i} + {\frac{1}{2}}Q_{B,i}-{\frac{1}{2}}Q_{S,i} \right) n_i , \nonumber \\
&=&n_I + {\frac{1}{2}}n_B^0 -{\frac{1}{2}}n_S,
\label{equation:n}\end{eqnarray}
where we have also used the definitions of the isospin density $n_I =  \Sigma_i Q_{I,i}n_i$ and strangeness density $n_S = \Sigma_i Q_{S,i}n_i$.
Dividing Eq.~\eqref{equation:n} by $n_B^0$ and using the definitions of isospin fraction $Y_I ={n_I}/{n_B^0}$ and strangeness fraction  $Y_S={n_S}/{n_B^0}$ results in:
\begin{eqnarray}
Y_Q =Y_I + \frac{1}{2} -\frac{1}{2}Y_S, 
\end{eqnarray}
so we can finally write:
\begin{eqnarray}
Y_I =Y_Q - \frac{1}{2} +\frac{1}{2}Y_S,
\label{vYi} 
\end{eqnarray} 
as a new way to calculate the isospin fraction in a formalism in which charge is the conserved quantity.

Combining  Eqs.~\eqref{equation:mui} and \eqref{equation:qi} gives:
\begin{eqnarray}
\mu_i &= &Q_{B,i} {\mu_B}+\left(Q_{I,i} + {\frac{1}{2}}Q_{B,i}-{\frac{1}{2}}Q_{S,i} \right)\mu_Q+ Q_{S,i}\mu_S, \nonumber \\
&= &Q_{B,i} \left(\mu_B + {\frac{1}{2}}\mu_Q \right) + Q_{I,i} {\mu_Q} + Q_{S,i} \left(\mu_S -\frac{1}{2}\mu_Q \right) ,\nonumber \\
&= &Q_{B,i} \ \mu_B' + Q_{I,i}\ {\mu_I} + Q_{S,i}\ \mu_S' .
\label{13}\end{eqnarray}
A comparison of the above with Eq.~\eqref{equation:mui} reveals that our conserved charge formalism is equivalent to another in which the isospin is fixed, leaving the isospin chemical potential as the independent chemical potential (together with $\mu_B$ and $\mu_S$), provided we use the following new variables:
\begin{eqnarray}
\mu_B' = \mu_B + {\frac{1}{2}}\mu_Q \ \ \ {\rm{and}} \ \ \ \mu_S' =  \mu_S - {\frac{1}{2}}\mu_Q .
\label{outra}
\end{eqnarray}
In this way, the chemical potentials correspond, $\mu_I = \mu_Q$, and  no modifications to our numerical codes are required to show isospin fractions and isospin chemical potentials. To the best of our knowledge, these model-independent fractions and isospin correspondences Eqs.~\ref{vYi}-\ref{outra} have never been discussed before. The expressions for the chemical potential of each particle included in the model derived using Eq.~\eqref{equation:mui} or Eq.~\eqref{13} are given in Appendix A. 

%***************** FIGURE 1: 3D YQ *************************************************************************************
\begin{figure*}[t!]
\centering
\begin{subfigure}[b]{0.475\textwidth}
\centering
\includegraphics[width=\textwidth]{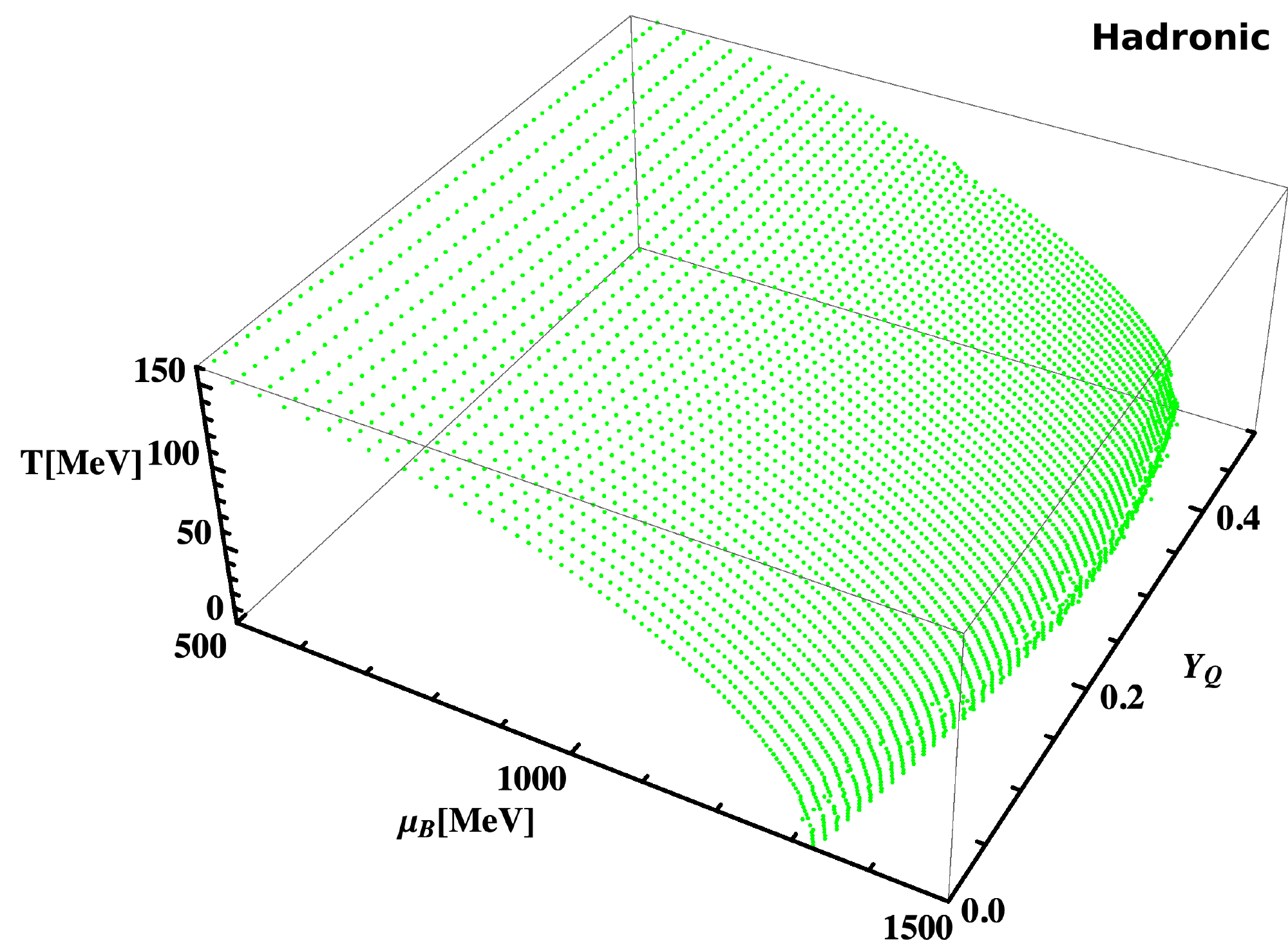}
  \end{subfigure}
 \quad
 \begin{subfigure}[b]{0.475\textwidth} 
\centering
 \includegraphics[width=\textwidth]{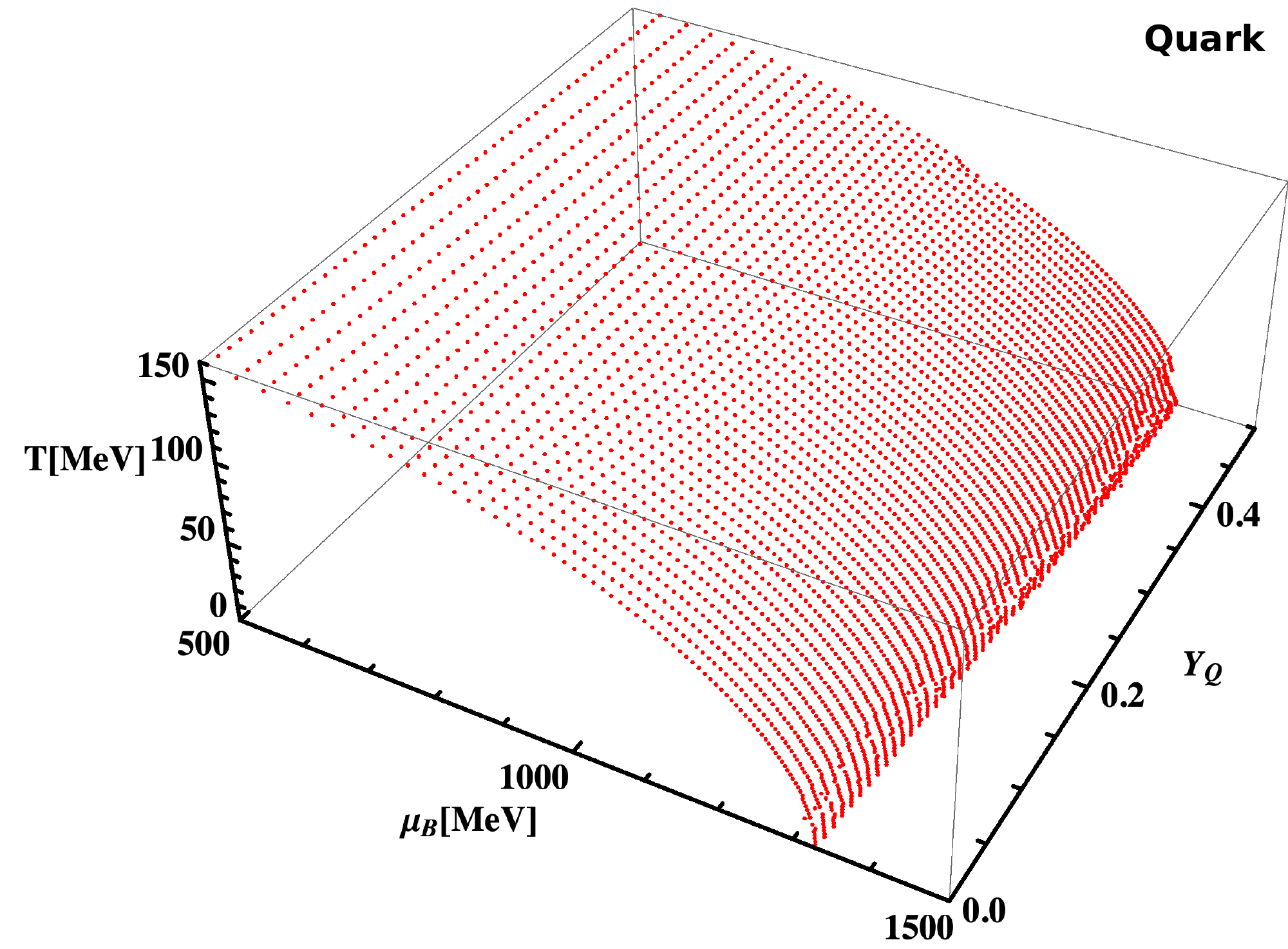}
  \end{subfigure}
\begin{subfigure}[b]{0.475\textwidth}  
\centering
\includegraphics[width=\textwidth]{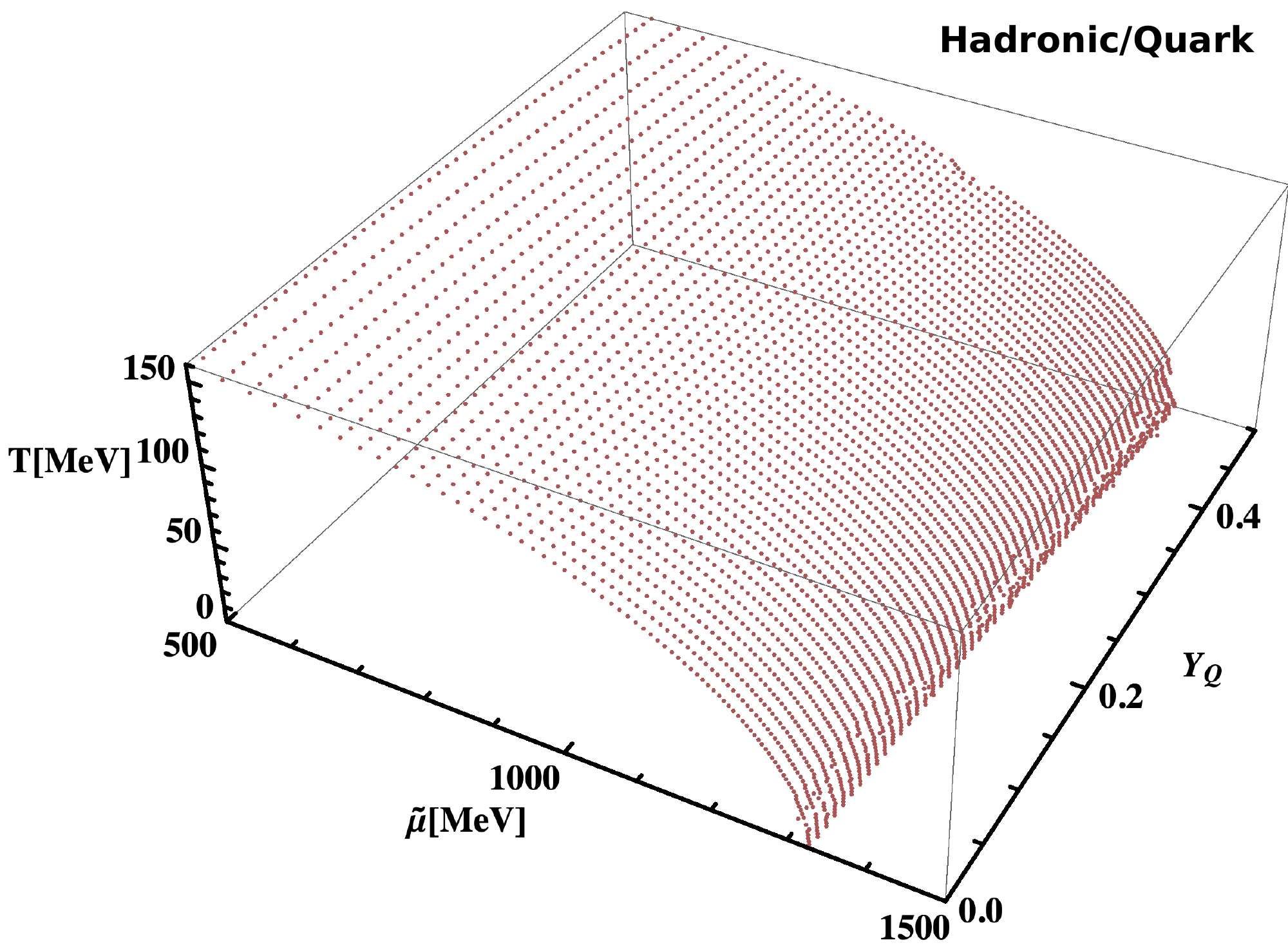}
  \end{subfigure}
\quad
  \caption {Top panels: The temperature T vs$.$ baryon chemical potential $\mu_B$ vs$.$ charge fraction $Y_Q$ phase diagram for non-strange matter $Y_S=0$ on the hadronic-phase side of the deconfinement phase transition coexistence region (left panel) and on the quark-phase side (right panel). Bottom panel: The temperature T vs$.$ free energy $\widetilde{\mu}$ vs$.$ charge fraction $Y_Q$ phase diagram for non-strange matter either on the hadronic or quark-phase side of the deconfinement phase transition coexistence region. All curves were calculated varying the charge fraction between $Y_Q=0$ and  $Y_Q=0.5$.}
\label{fig:fig}
\end{figure*}  
%******************************************************************************************************************

It is also convenient to define a Gibbs free energy per baryon (henceforth called simply free energy) of the system, a quantity that is always the same on both sides of a first-order phase transition in order to fulfill phase equilibrium. In our case (when, besides baryon number, charge fraction and strangeness fraction are also fixed), it is: 
\begin{eqnarray}
&\widetilde{\mu} &= \mu_B + Y_Q\mu_Q + Y_S\mu_S .
\label{ve2}
\end{eqnarray}
Note that the free energy will be equal to the baryon chemical potential only in the particular cases of zero charge fraction or zero charge chemical potential and zero strange fraction or strange chemical potential. This is the case in the modeling of the typical examples of deleptonized cold neutron stars (charge neutral in chemical equilibrium $Y_{Q_{\rm{total}}}=0$, $\mu_Q=-\mu_e$ and with no constraint on net strangeness $\mu_S=0$) and relativistic HICs (no net isospin $\mu_Q=0$ and no net strangeness $Y_S=0$). 

%***************** FIGURE 2: 3D muQ*************************************************************************************
\begin{figure*}[t!]
\centering
\begin{subfigure}[b]{0.475\textwidth}
\centering
\includegraphics[width=\textwidth]{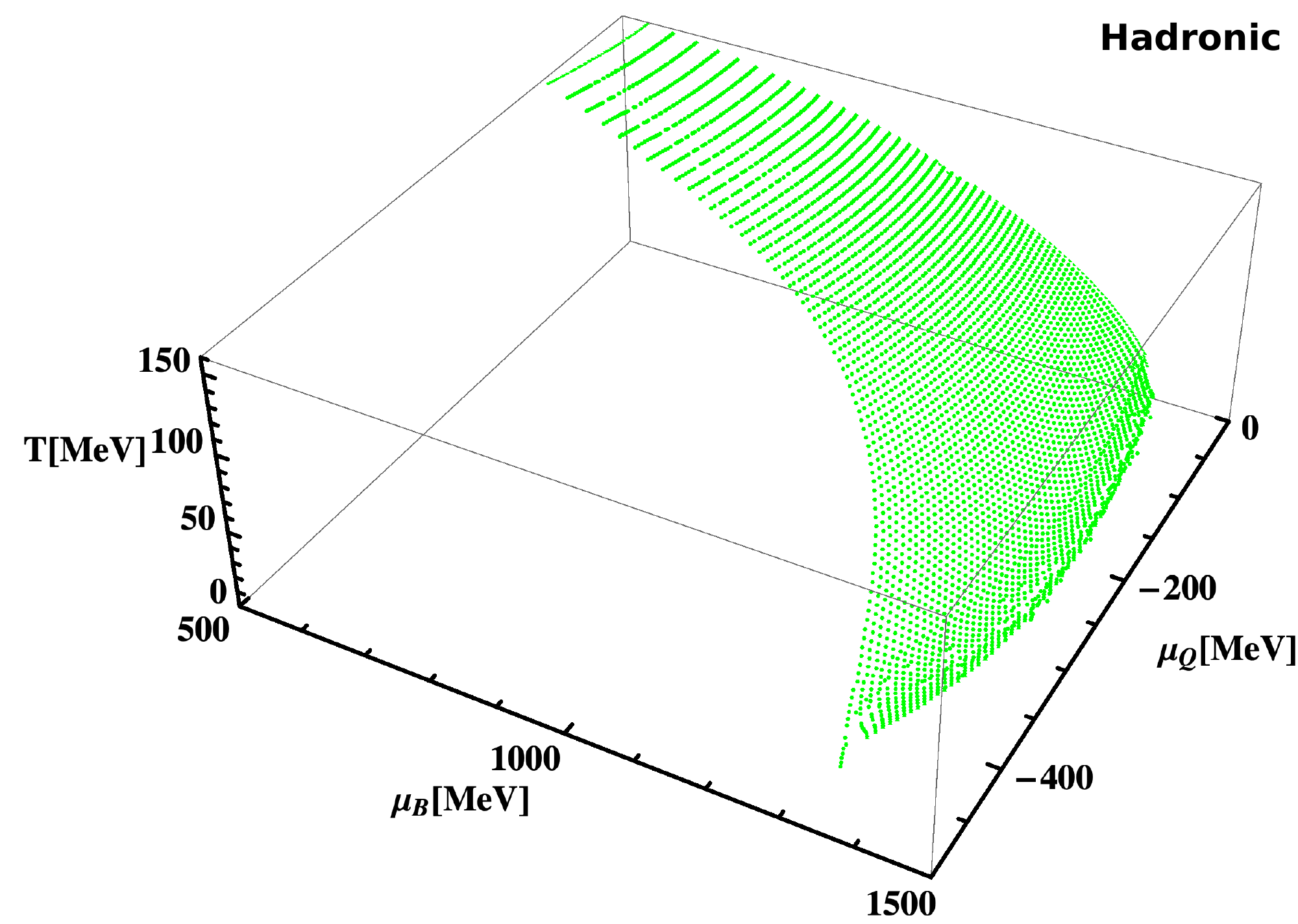}
 \end{subfigure}
 \quad
 \begin{subfigure}[b]{0.475\textwidth} 
\centering
 \includegraphics[width=\textwidth]{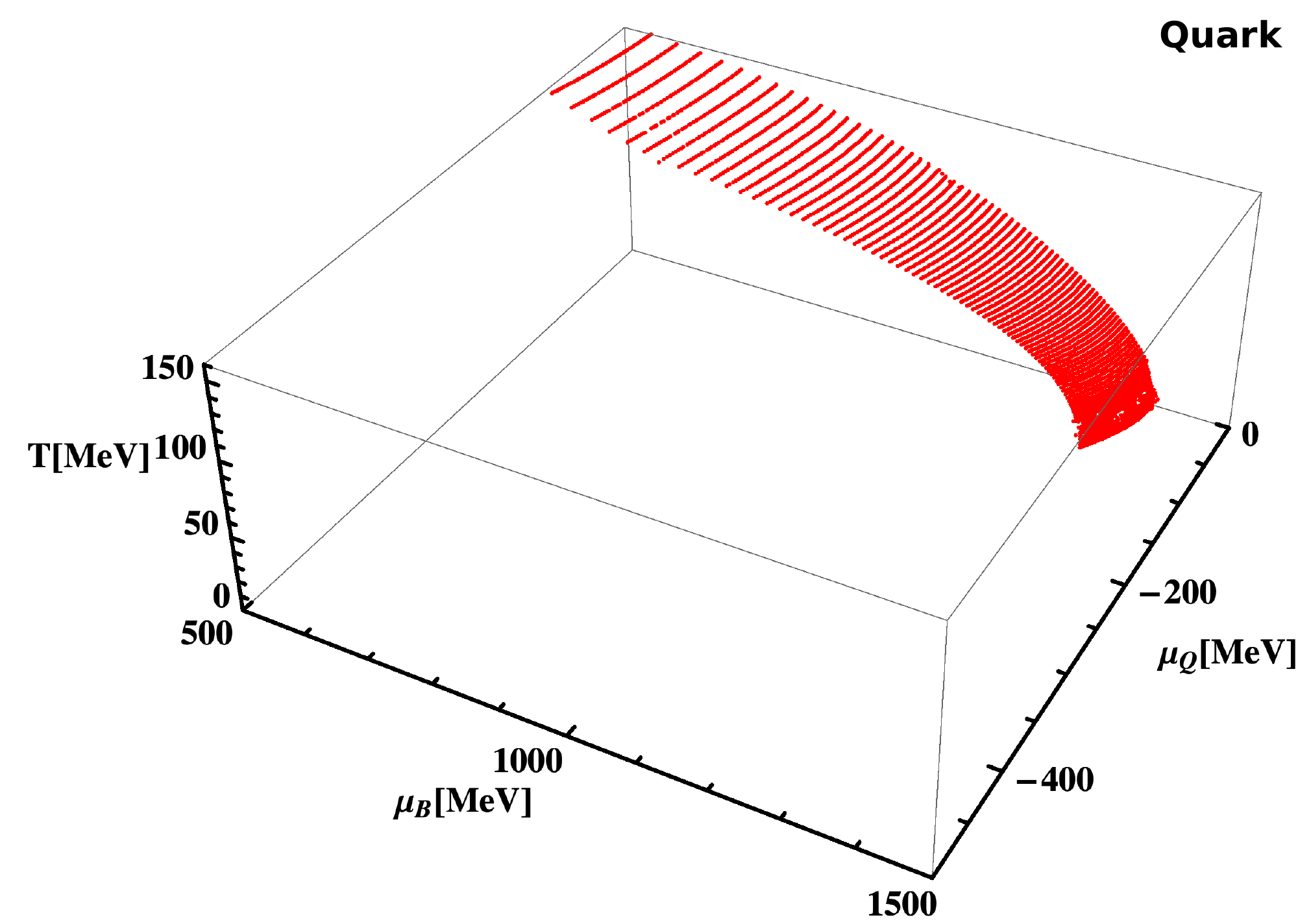}
  \end{subfigure}
\begin{subfigure}[b]{0.475\textwidth}  
\centering
\includegraphics[width=\textwidth]{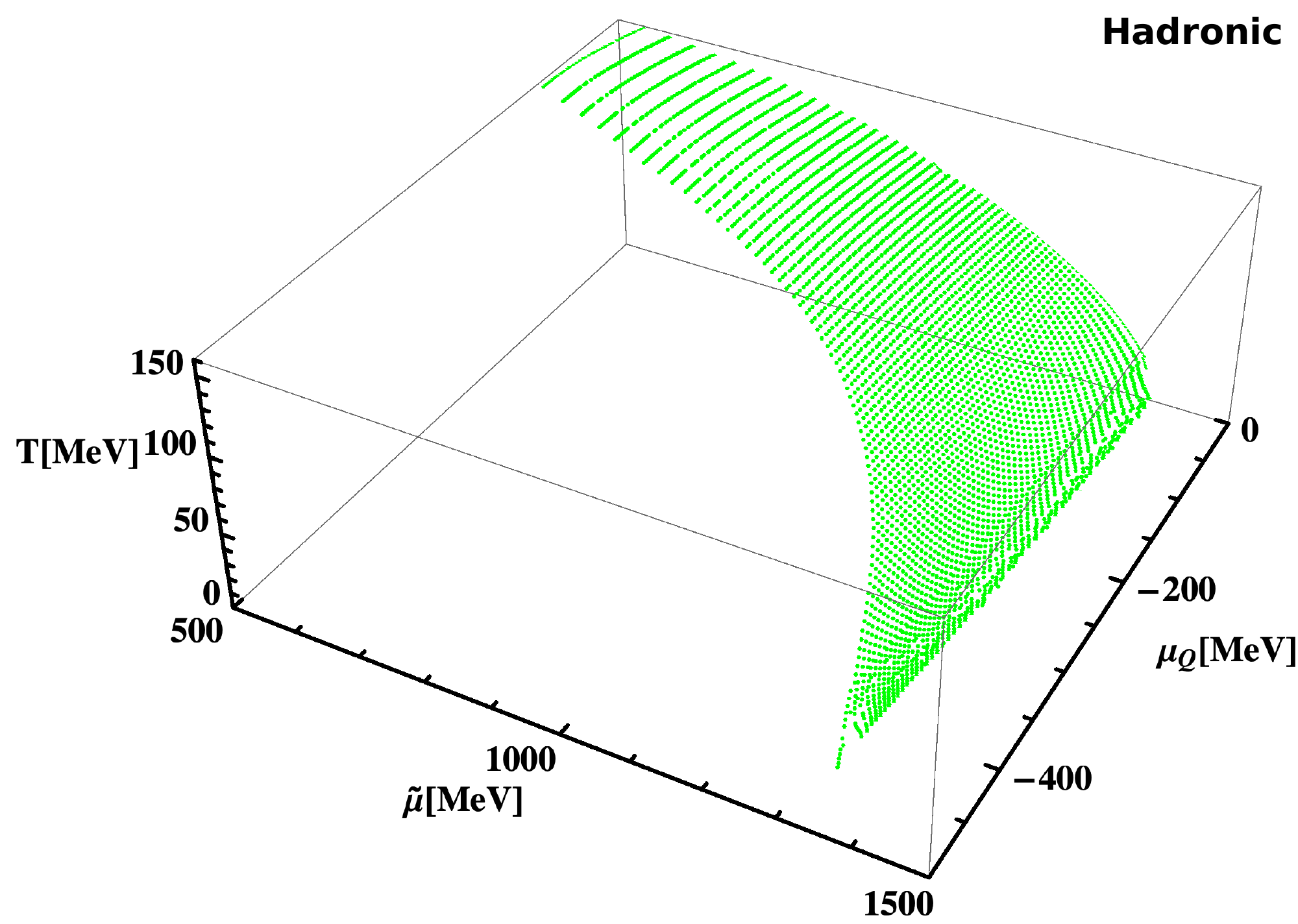}
  \end{subfigure}
\quad
\begin{subfigure}[b]{0.475\textwidth}  
\centering
\includegraphics[width=\textwidth]{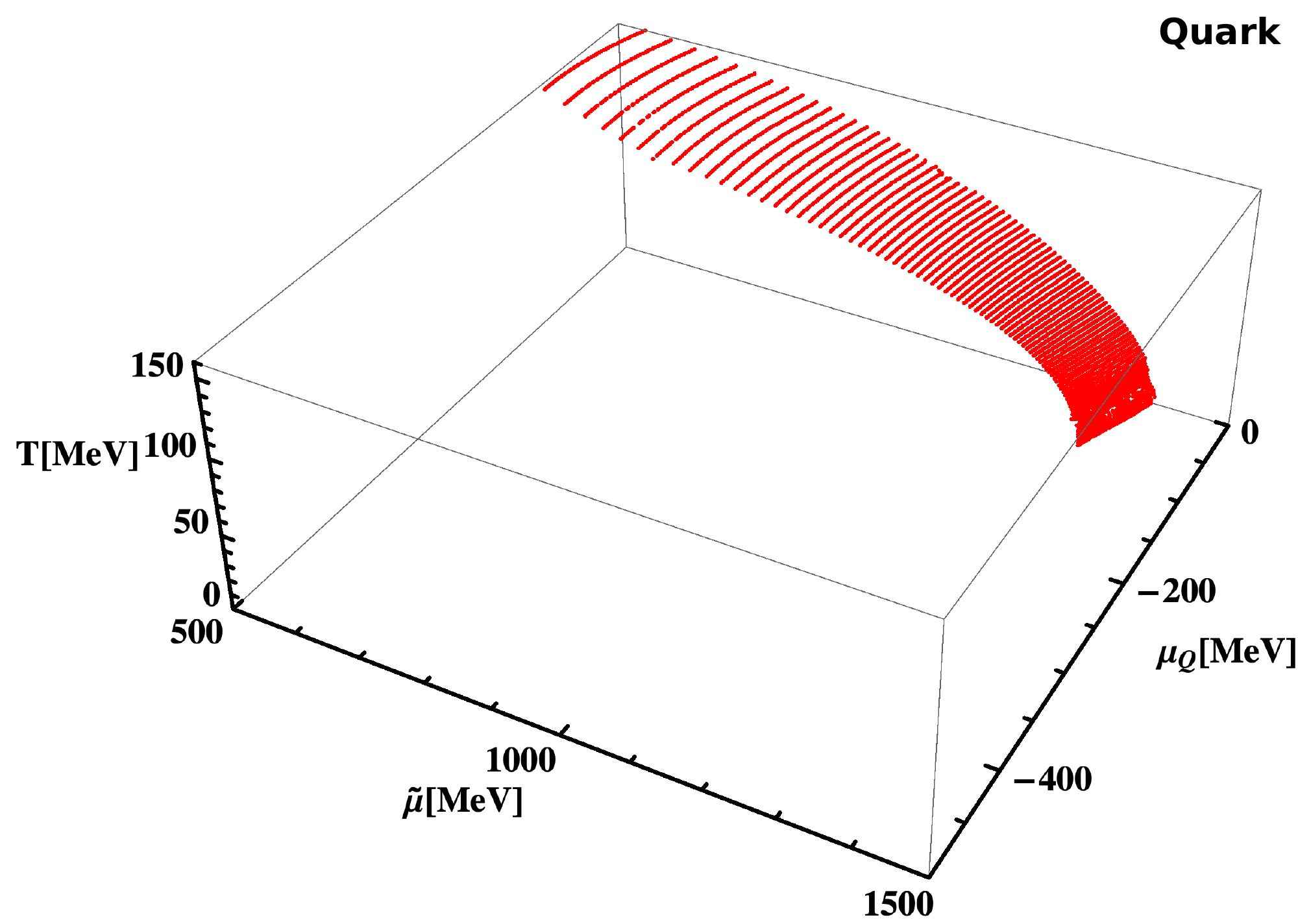}
  \end{subfigure}
   \caption {Same as Fig.~\ref{fig:fig} but showing the charged chemical potential $\mu_Q$. The separate bottom panels show the hadronic phase (left panel) and quark phase (right panel) sides of the deconfinement phase transition coexistence region.}
\label{fig2:fig}
\end{figure*} 
%******************************************************************************************************************

Eq.~\eqref{ve2} was derived without any model assumption throughout Ref.~\cite{Hempel_2009}  for the particular case in which net strangeness is not constrained (which implies $\mu_S=0$) and charge neutrality is enforced by leptons. It was then later discussed in full in the Appendix D of Ref.~\cite{Hempel:2013tfa}. This equation can be derived either by a Legendre transformation of the grand-potential or by taking the derivative of minus the grand-potential with respect to the baryon number, giving:
\begin{eqnarray}
&\widetilde{\mu}={\sum_i\mu_in_i}/{n_B^0} .
\label{ve}
\end{eqnarray}
Substituting $\mu_i$ from Eq.~\eqref{equation:mui} in Eq.~\eqref{ve} results in:
\begin{eqnarray}
&\widetilde{\mu}=\frac{\big(\sum_i Q_{B,i} n_i \big) \mu_B}{n_B^0}  + \frac{\big(\sum_i Q_i n_i \big) \mu_Q}{n_B^0}  +  \frac{\big(\sum_i Q_{S,i} n_i \big) \mu_S}{n_B^0} .\ \ \ 
\label{ve'}
\end{eqnarray}
which, using the definitions above, results in Eq.~\eqref{ve2}. Alternatively, replacing $\mu_i$ from Eq.~\eqref{13} in Eq.~\eqref{ve} leads to:
\begin{eqnarray}
&\widetilde{\mu} &= \mu_B' + Y_I\mu_I + Y_S\mu_S' .
\label{lst}
\end{eqnarray}
This general equation for the free energy in a formalism in which the isospin and strangeness are conserved had not been discussed before in the literature.

In the particular case of the CMF model, the grand-potential density of the system is:
\begin{eqnarray}
\Omega=-P = \varepsilon - T s - \sum_i \mu_i n_i - \mu_B n_\Phi ,
\label{omega1}
\end{eqnarray}
where the last term is a non-baryonic contribution $n_\Phi = -(4 a_1 \mu_B^3 + 2 a_2 T^2 \mu_B) \Phi^2$ (with negative coefficients) necessary to ensure thermodynamical consistency. This follows from the fact that our potential $U$ for $\Phi$ contains baryon chemical potential terms and, therefore, contributes to $n_B$. One can understand these terms as arising from a chemical-potential dependence of the confinement phase transition temperature, see e.g. discussions in Refs.~\cite{Schaefer:2007pw,Fukushima:2010pp,Shao:2016fsh}.

Replacing Eq.~\eqref{equation:mui} in \eqref{omega1} gives:
\begin{eqnarray}
\Omega&=&\varepsilon - T s - \Big(\sum_i Q_{B,i} n_i \Big) \mu_B \\ \nonumber
 &-& \Big(\sum_i Q_i n_i \Big) \mu_Q -  \Big(\sum_i Q_{S,i} n_i \Big) \mu_S - \mu_B n_\Phi ,
\label{omega2}
\end{eqnarray}
\begin{eqnarray}
\Omega&=&\varepsilon - T s - {n_B^0} \mu_B 
 - n_B^0 Y_Q \mu_Q -  n_B^0 Y_S \mu_S - \mu_B n_\Phi .\nonumber\\
\label{omega3}
\end{eqnarray}
Taking the derivative of minus the CMF grand-potential density Eq.~\ref{omega3} with respect to the baryon number divided by volume $n_B^0$ gives once more Eq.~\eqref{ve2}.

\section{Results}
\subsection{Non-Strange Matter $Y_S=0$}

We start by discussing 3-dimensional phase diagrams with first-order phase transition deconfinement coexistence regions calculated within  the CMF model for temperatures in the range $0-160$ MeV, charge fractions in the range $0-0.5$, and the corresponding baryon chemical potentials $\mu_B$ or free energies $\widetilde{\mu}$. In order to construct those, at each given temperature and charge fraction, the free energy is varied. The free energy of the deconfinement coexistence region is determined by finding a jump in the deconfinement order parameter $\Phi$. This jump is very large (basically from $0\to1$) at zero temperature, but its size decreases with temperature until it becomes very close to zero at our chosen maximum temperature near the critical point. At the coexistence region, our numerical code determines the baryon chemical potential (see equation below) and charged chemical potential that reproduce the given charge fraction. In addition, in this subsection, the strange chemical potential is also determined numerically in order to produce a zero net strangeness $Y_S=0$ in each phase. This is the case for matter produced in HICs, where there is no time for strangeness to emerge. Note that for non-strange matter Eq.~\eqref{ve2} simplifies to:
\begin{eqnarray}
&\widetilde{\mu} &= \mu_B + Y_Q\mu_Q .
\label{ve29}
\end{eqnarray}

Having $Y_Q = 0$ means that there is no net charge in the system even though the presence of charged particles is not prohibited insofar as the sum of their charges is zero. Having $Y_Q = 0.5$ corresponds to the situation where the total baryon number of the system is twice as large as its net charge. For matter with no net strangeness at zero temperature, the case of $Y_Q = 0$ is equivalent to having just neutrons or two times more d-quarks than u-quarks, whereas $Y_Q = 0.5$ corresponds to having equal amounts of protons and neutrons or d- and u-quarks. At finite temperature, there can be hyperons and s-quarks present when requiring no net strangeness, as long as the difference between the number of strange particles and strange antiparticles is zero. 

As shown in the bottom panel of Fig.~\ref{fig:fig}, the free energy at deconfinement increases as a function of $Y_Q$. This behavior is related to the softening of nuclear matter with increased net charge (e.g. equal numbers of neutrons and protons), the effect being stronger for hadronic matter. A softening of the equation of state (pressure vs. energy density) of hadronic matter corresponds to an increase in pressure at a given free energy (with respect to the quark phase), therefore, extending the stability of the hadronic phase to larger free energies.

%***************** FIGURE 3: 2D YQ/muQ*************************************************************************************
\begin{figure*}[t!]
\centering
\begin{subfigure}[b]{0.49\textwidth}
\centering
\includegraphics[trim={.89cm 0 2.35cm 0},clip,width=.99\textwidth]{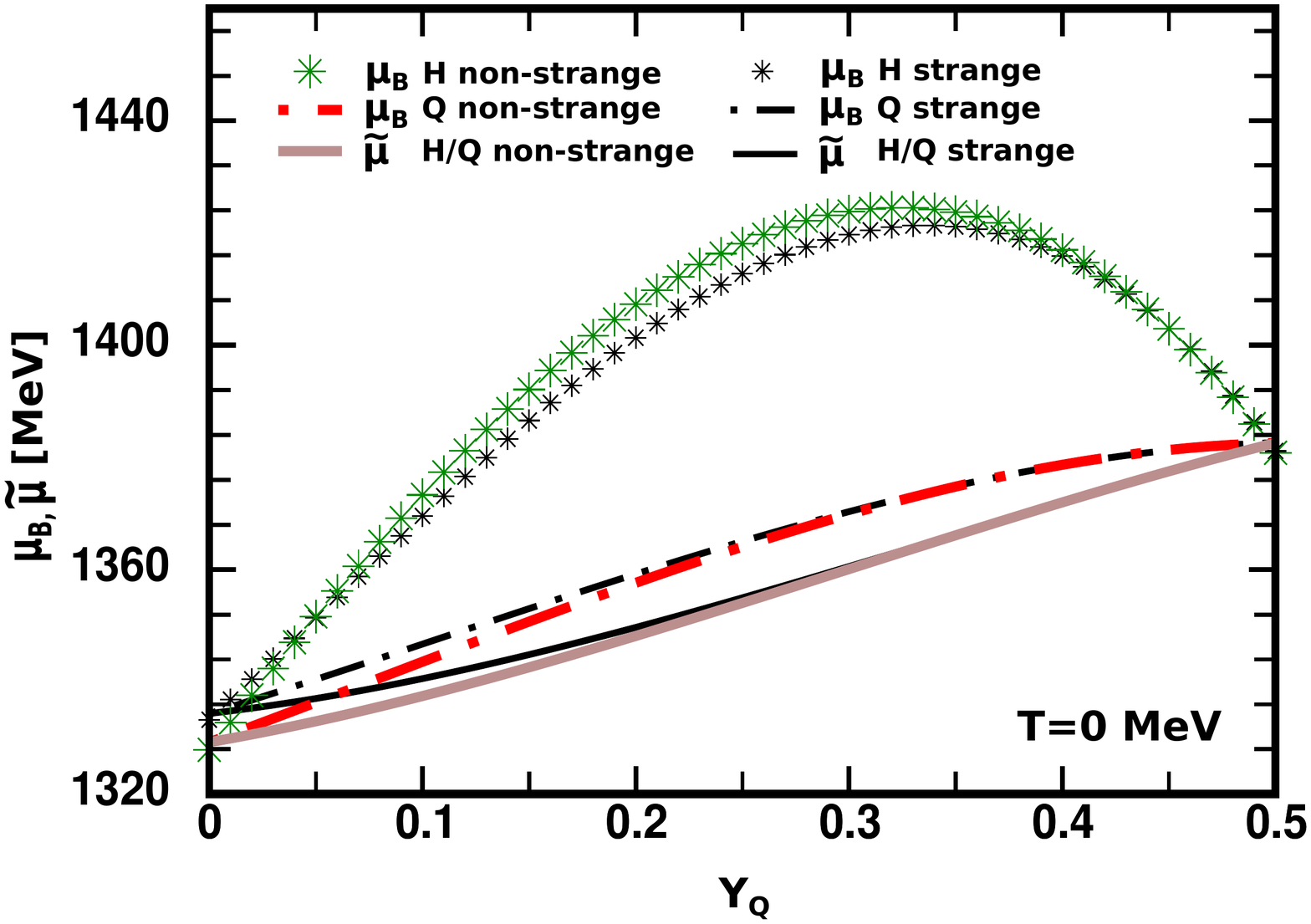}
  \end{subfigure}
 \begin{subfigure}[b]{0.49\textwidth} 
\centering
 \includegraphics[trim={0.42cm 0.08cm 2.49cm 0},clip,width=1.\textwidth]{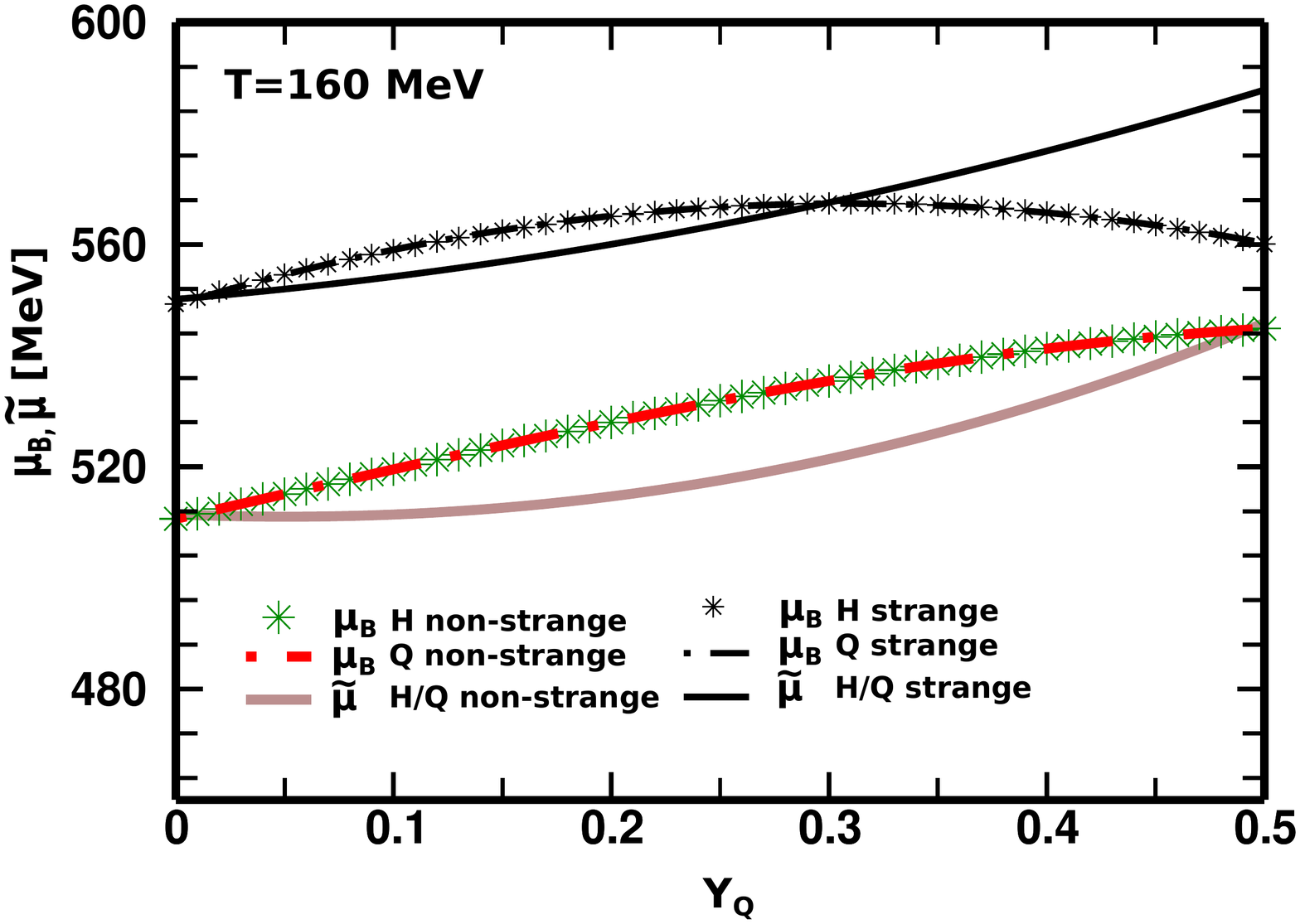}
  \end{subfigure}
\centering
\begin{subfigure}[b]{0.475\textwidth}
\centering
\includegraphics[trim={.97cm 0 2.6cm 0},clip,width=\textwidth,width=.97\textwidth]{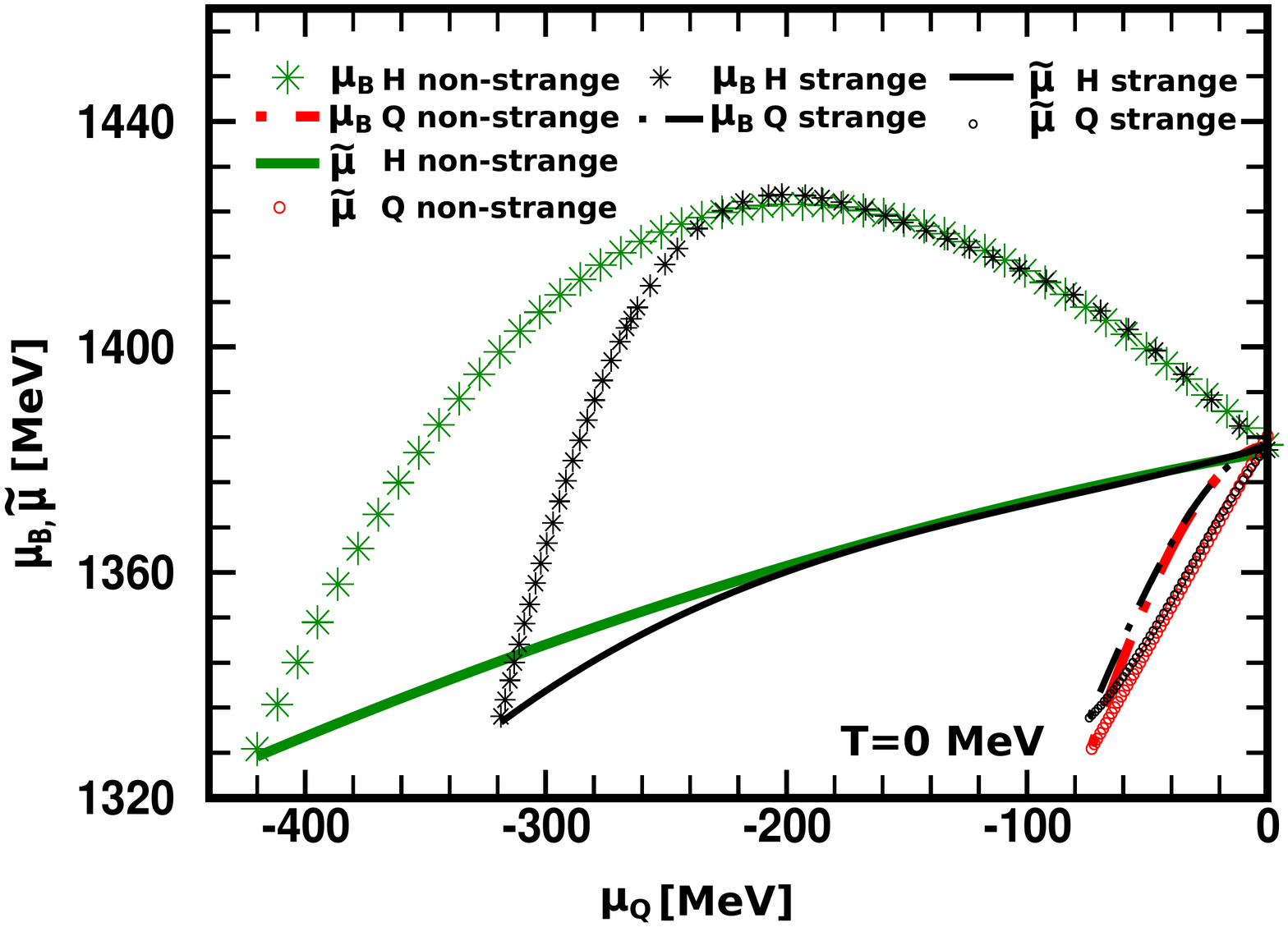}
 \end{subfigure}
 \begin{subfigure}[b]{0.475\textwidth} 
\centering
 \includegraphics[trim={.8cm 0.6cm 2.26cm 0},clip,width=1.03\textwidth]{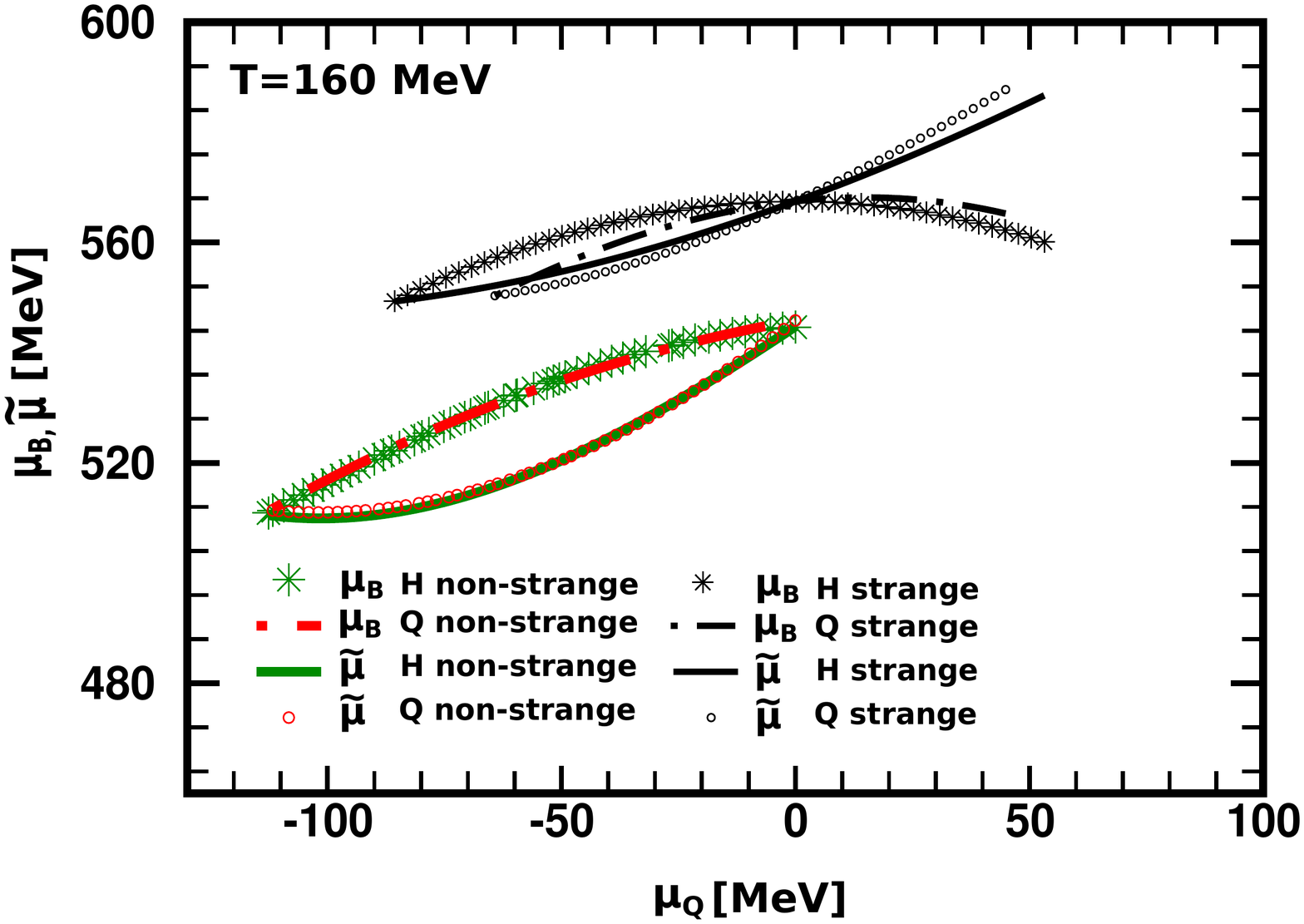}
  \end{subfigure}
\caption {Equivalent to Fig.~\ref{fig:fig} (top panels) and Fig.~\ref{fig2:fig} (bottom panels) but only showing results along the deconfinement coexistence line for $0$ (left panels) and $160$ MeV (right panels) temperatures. Green, red and brown lines (all grey in black and white print) show results already discussed for non-strange matter, while black lines show new results for strange matter.} 
\label{fig6:fig}
\end{figure*}  
%******************************************************************************************************************

Although the free energy is the same on both sides of the deconfinement coexistence region, the baryon chemical potential is not. This is evident in a comparison of the top left (hadronic phase) to the top right (quark phase) panel of Fig.~1. The difference stems from the fact that the baryon chemical potential is calculated from the free energy using the charged chemical potential, which is different on either side of the phase transition (the reason for which will be discussed in the following). In addition, when comparing the top left panel with the bottom one, we find a reasonable difference for all cases corresponding to $\mu_B \neq \widetilde{\mu}$ in Eq.~\eqref{ve29}, that is, for all $Y_Q$ other than 0 and 0.5 (when $\mu_Q=0$).

The difference is much smaller between the top right panel of Fig.~\ref{fig:fig} and the bottom one, as the charged chemical potential $\mu_Q$ is always small in the quark phase. This has already been shown in Fig.~3 of Ref.~\cite{Roark:2018uls} for the particular case of chemically equilibrated matter (with and without trapped neutrinos). Here, we extend this discussion to matter out of chemical equilibrium. A comparison of the left and right panels of Fig.~\ref{fig2:fig} demonstrates that the hadronic-phase side reaches much larger absolute values of $\mu_Q$ than the quark phase for small charge fractions (corresponding to the more negative $\mu_Q$'s). This can be easily understood in the case of zero temperature. In this case, $Y_Q = 0$ means having only neutrons, which requires a very large difference between their chemical potential and the proton one (that differ only by $\mu_Q$, as shown in the equations of Appendix A). In the quark phase at zero temperature, $Y_Q = 0$ implies having twice the amount of d-quarks than u-quarks, a much more balanced case that requires a smaller $\mu_i$ difference and, therefore, a smaller $\mu_Q$ absolute value.

%*****************FIGURE 4: 2D YI*************************************************************************************
\begin{figure*}[t!]
\centering
\begin{subfigure}[b]{0.475\textwidth}
\centering
\includegraphics[trim={.8cm 0 2.1cm 0},clip,width=0.99\textwidth]{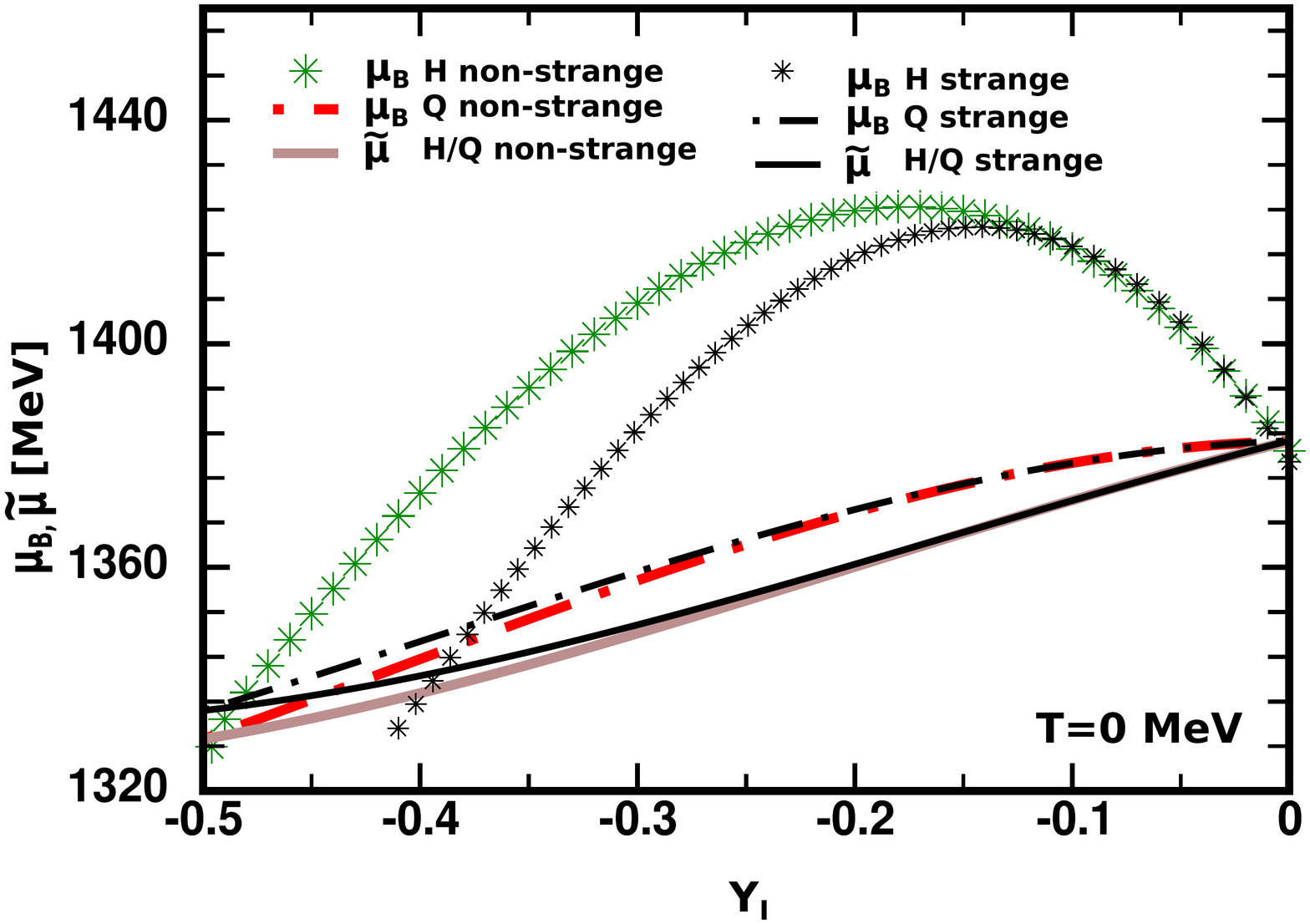}
  \end{subfigure}
 \begin{subfigure}[b]{0.475\textwidth} 
\centering
 \includegraphics[trim={1.2cm .25cm 1.63cm 0},clip,width=1.\textwidth]{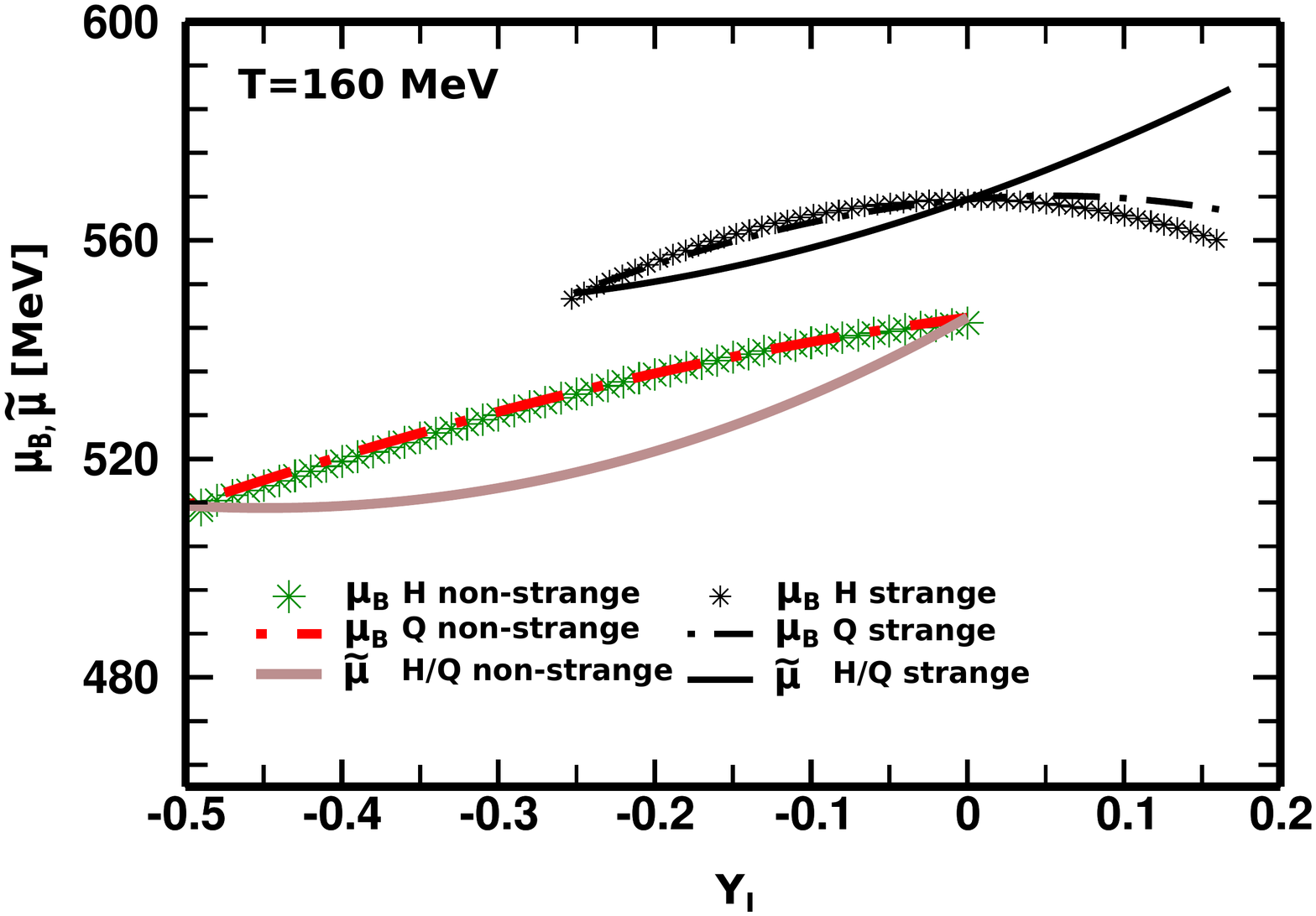}
  \end{subfigure}
  \caption {Same as the top panels of Fig.~\ref{fig6:fig} but showing the isospin fraction $Y_I$. Equivalent isospin chemical potential panels would be exactly like the charged chemical potential bottom panels of Fig.~\ref{fig6:fig}.}
\label{fig8:fig}
\end{figure*}
%******************************************************************************************************************

In the case of Fig.~\ref{fig2:fig} (unlike Fig.~\ref{fig:fig}), the bottom panels are always different from each other. This happens because the charged chemical potential itself is discontinuous across the first-order phase transition. Analyzing separately the hadronic-phase (left) side of the coexistence region in Fig.~\ref{fig2:fig}, it can be seen that the curves in the top and bottom panels are always different, except on the upper and lower boundaries of $\mu_Q$.

We show phase diagrams as functions of charge fraction because this is common practice in astrophysics, where the requirement of charge neutrality implies $Y_Q = Y_{\rm{electron}}$. There is no  corresponding general equality for the electron chemical potential: the relation $\mu_Q=-\mu_{\rm{electron}}$ is only valid in the special case of chemical equilibrium, which is only established in deleptonized cold neutron stars. Similar figures to Figs.~\ref{fig:fig} and \ref{fig2:fig} are presented in Appendix B for the equivalent scenario of fixed isospin fraction. For non-strange matter, Eq.~\eqref{vYi} reduces simply to $Y_I=Y_Q-0.5$ and, therefore, the changes in both figures are trivial. More details are given in Appendix B.

\subsection{Strange matter $Y_S\neq0$}
  
In this subsection, we compactify the temperature and only show results for $T=0$ MeV and $T=160$ MeV (corresponding to the two temperature extremes in our previous figures) in order to make comparisons. Full 3-dimensional phase diagrams with net strangeness are available upon request. In addition to quantities shown in the preceding subsection (for matter with net strangeness constrained to zero) using the same colors, we now display strange matter results in black for comparison. By strange matter, we mean matter in which there is no constraint on net strangeness and therefore, no strange chemical potential ($\mu_S=0)$. For $T=0$, no significant difference in the position of the deconfinement phase-transition coexistence line with respect to the baryon chemical potential or free energy is expected due to allowing for nonzero strangeness. This is because, in this case, in our model, only a few $\Lambda$'s and $\Sigma^-$'s are present around the deconfinement free energy (and no strange quarks) and only at small charge fractions. This is illustrated in the difference between the solid brown line and the solid black line in the upper left panel of Fig.~\ref{fig6:fig}. The strange solid black line being at a larger $\widetilde{\mu}$ than the non-strange brown dashed line is a consequence of the hyperons softening hadronic matter when they appear (for low charge fraction) and pushing the phase transition coexistence line to larger values of $\widetilde{\mu}$.

The difference due to strangeness in the position of the coexistence line with respect to the baryon chemical potential is related to the presence of strange particles, which modify the charged chemical potential relative to the zero-strangeness case (Eq.~\eqref{ve2} reduces once more to  Eq.~\eqref{ve29} for $\mu_S=0$). As a consequence, as shown in the green vs. black stars still in the upper left panel of Fig.~\ref{fig6:fig}, $\mu_B$ on the hadronic-phase side is lower around intermediate charge fractions for the strange case. This is a combination of  $\mu_Q$ being lower in absolute value for low and intermediate values of $Y_Q$ for strange matter (see different color stars in the bottom left panel of Fig.~\ref{fig6:fig}) and the fact that $\mu_Q$ is multiplied by $Y_Q$ in Eq.~\eqref{ve29}. As for the quark-phase side of the phase transition coexistence line, $\mu_Q$ is always small in absolute value (different color dotted-dashed lines in bottom left panel of Fig.~\ref{fig6:fig}), so $\mu_B$ behaves very similarly to $\widetilde{\mu}$ in strange (as well as in non-strange matter), as seen when comparing black and red dot-dashed and solid lines in the top left panel of Fig.~\ref{fig6:fig}.

For large temperatures, strangeness generates much larger effects and the first-order phase transition itself is very weak (particularly for the case without net strangeness), becoming very similar to a crossover. The large effects translate into a significant difference in the position of the black vs. colored lines in the top right panel of Fig.~\ref{fig6:fig}: the strange black solid line for $\widetilde{\mu}$ resides about $40$ MeV higher than the non-strange dashed pink one. For T=0, this difference is $\lesssim5$ MeV. This large shift is a consequence of the fact that, at large temperatures, the presence of strangeness-carrying particles is enhanced at all charge fractions, but now softening more the quark equation of state (relative to the hadronic one) around deconfinement. As a consequence, strangeness pushes the free energy to larger values.

To discuss the baryon chemical potential, we first note that at this large temperature, which is very close to the critical point, the hadronic-phase and quark-phase sides of the deconfinement phase transition are nearly identical. The difference in the position of  $\mu_B$ with respect to $\widetilde{\mu}$ has to do with the fact that, once again, the charged chemical potential difference also needs to be accounted for. When looking at the black stars and dot-dashed line still in the upper right panel of Fig.~\ref{fig6:fig}, we find that they are lower in comparison to the solid line for $\widetilde{\mu}$ (than in the colored non-strange case). This has to do with the fact that $\mu_Q$ is lower in absolute value and even positive for some large charge fractions when strangeness is included (see all black lines in the right bottom panel of Fig.~\ref{fig6:fig}).

Fig.~\ref{fig8:fig} shows the effects of strangeness on the baryon chemical potential and free energy as a function of the isospin fraction $Y_I$. Now, when net strangeness is non- zero (black curves), the left panel in this figure is not simply a constant horizontal shift from the $Y_Q$ axis shown in the top left panel of the previous figure, but a shift that, according to Eq.~\eqref{vYi}, depends on the strangeness fraction and, therefore, is different for every point. The horizontal shift is always positive and larger for low $Y_I$/$Y_Q$ at zero temperature, where there is more net strangeness. At $T=160$ MeV, the black lines in the right panel of Fig.~\ref{fig8:fig} show that the horizontal shift (with respect to the upper-right panel of the previous figure) is always positive and substantial for all $Y_I$/$Y_Q$, as, in this case, net strangeness is always present.

Note that in Fig.~\ref{fig8:fig} we do not show bottom panels for isospin chemical potential, as they would be identical to the charged chemical potential bottom panels of Fig.~\ref{fig6:fig}. In addition, if instead of using Eq.~\eqref{vYi} to calculate $Y_I$, we had rewritten our numerical code to run for fixed isospin fraction from -0.5 to 0, we would have obtained the same results as shown in Fig.~\ref{fig8:fig} but with an extra piece on the left and a missing piece on the right side of our finite temperature panel, a consequence again of the non-trivial $Y_Q$ to $Y_I$ shift.

\section{Discussion and Conclusions}

We present, for the first time, a comprehensive study of the effects of fixing and varying either the (hadronic and quark) charge fraction or isospin fraction on the position of the deconfinement to quark matter coexistence line. To do so, we assume the deconfinement phase transition to be of first order and make use of the Chiral Mean Field (CMF) model to produce our equations of state. We start by obtaining model-independent relations among charge and isospin fractions including how they are affected by the presence of net strangeness. We also show the relation between the isospin and charge chemical potentials and free energies. We then use these relations to draw 3-dimensional high-energy phase diagrams showing phase-transition coexistence regions for the CMF model. This discussion is extremely timely as, historically, the heavy-ion collision community has modeled their systems in terms of fixed isospin fraction, while the astrophysical community has modeled it in terms of charge fraction (equal to the electron fraction when muons are not included), whereas now these communities are working together to understand the hot and dense matter generated in neutron star mergers and in low energy heavy-ion collisions and need to have their findings compared. We provided here this tool.

Our goal in this work has been to obtain a qualitative description of how a given fixed charge fraction or isospin fraction changes the position of the deconfinement coexistence line (for a given temperature) to larger or lower baryon chemical potential or Gibbs free energy per baryon. To that end, we have built 3-dimensional phase diagrams for matter that possesses no net strangeness, the kind of matter created in particle colliders like RHIC and LHC. We have also determined the ranges that can be probed (given specific initial conditions of temperature and strangeness) for charge and isospin chemical potentials during deconfinement along the phase-transition coexistence line. Unlike quark matter produced in the lab, quark matter created inside stars can be strange, as the timeframe for its creation is much longer than the timeframe for weak decay. To discuss the effects of net strangeness on deconfinement to quark matter, we have constructed 2-dimensional phase diagrams at two chosen temperatures of $T=0$ and $T=160$ MeV. In the former case, very little strangeness is created and, therefore, its effects are minimal. In the latter, the consequences of nonzero strangeness are significant. 

For example,  when the charge fraction changes from $Y_Q=0\to 0.5$, the baryon chemical potential at the deconfinement coexistence line can change by up to $130$ MeV (at zero temperature on the hadronic side), the free energy by up to $50$ MeV (at zero temperature), and the charge/isospin chemical potential by up to $330$ MeV (at zero temperature on the hadronic side). At zero temperature, we have found that, for non-strange matter ($Y_S=0$), the charged and isospin chemical potentials $\mu_Q$  and $\mu_I$ cover a range from $-420$ to $0$ MeV following the deconfinement coexistence line, reaching more negative values on the hadronic-phase side of the phase transition. For the strange case $Y_S\neq0$, the corresponding range is $-320$ to $0$ MeV, once again reaching more negative values on the hadronic-phase side of the phase transition. On the quark-phase side of the phase transition, $\mu_Q$  and $\mu_I$ lie between $-75$ and 0 MeV. At large temperatures close to the critical point, $\mu_Q$  and $\mu_I$ become practically  the same on the hadronic and the quark sides and have intermediate values for $Y_S=0$ ranging from $-110$ to $0$ MeV. Finally, when strangeness is allowed, $\mu_Q$  and $\mu_I$ at large temperature become less negative and even positive, reaching $\sim50$ MeV (all values calculated following the deconfinement coexistence line).

Our results show that comparisons among results from heavy-ion collision and hot astrophysical scenarios concerning the position of the deconfinement phase transition have to be interpreted carefully. Their different characteristics i.e in charge fraction going from $Y_Q\sim0.4-0.5$ to $Y_Q\sim0.1-0.15$ can change considerably (by hundreds of MeV in chemical potentials for a given temperature) the position of the deconfinement phase-transition coexistence line. Also, when strangeness is included, the commonly discussed equivalence between $Y_Q=0.5$ and $Y_I=0$ is broken and in reality correspond to very different systems.

Note that the formulas presented in section IIB, concerning the relation between charge and isospin fractions and respective chemical potentials are independent of any chosen microscopic model to describe different phases. Concerning our quantitative results extracted from phase diagrams, they are model dependent, as the equation of state in each phase depends on the particle population included and particle couplings, both of which are hard to quantify.  Nevertheless, at zero temperature, the amount of hyperons and strange quarks allowed in models used to describe neutron stars, as well as how isospin/charge fractions modify the equation of state are to some extent constrained by several laboratory and astrophysical observations, among which we list the symmetry energy, its slope, and the hyperon optical potentials.

More specifically, on one hand, hyperons cannot appear at very low densities due to their massive character and (together with protons) cannot appear in a very large number, while keeping matter from being too soft (and unable to generate massive stars) and preventing stars from cooling too fast \cite{Dexheimer:2015qha}.  On the other hand, there are no indications that quarks appear close to saturation for isospin symmetric matter, but they also cannot appear too late, beyond densities in which hadrons are expected to overlap. Finally, the strange quarks usually appear at larger densities than the light ones. The uncertainty  included in our calculations at finite temperature is larger, as it becomes harder to find observables to fit effective models at large densities. In this case, we rely on comparisons with perturbative QCD to test our model \cite{Roark:2018uls}. That being said, our results concerning the dependence of the position of the deconfinement phase transition on charge/isospin fraction is only due to the stiffening/softening of the hadronic/quark equations of state when different particles appear and Fermi levels are occupied by a different amount, which strongly depends on the particles' quantum numbers. As a result of the fractional nature of the quark quantum numbers, quark matter is not as sensitive to (small) $Y_Q$ as hadronic matter. As a consequence, quark matter does not respond as much for a given change in $Y_Q$. In this sense, the effective model used to produce our results is only a tool fitted to reproduce a reasonable particle population at each density/temperature/charge or isospin fraction.

As a final note, it is known that systems that undergo a first-order phase transition between phases in which more than one charge is globally conserved are non-congruent and present extended mixtures of phases. These cases show no discontinuities in the first derivatives of the potential \cite{Glendenning:1992vb}. An example is the case we discuss in this work, where charge or isospin is conserved in addition to baryon number. An exception takes place when the charge or isospin fractions are $Y_Q=0.5$ or $Y_I=0$, implying that the respective chemical potentials are zero (for an extended discussion on this azeotropic behavior, see Section II of Ref.~\cite{Hempel:2013tfa}). Another exception takes place when the surface tension between the phases is too large and electric charge is conserved locally instead of globally. For this work, we assume the latter and, therefore, avoid the discussion of a mixture of phases. But, even if that had not been the case, and we had chosen to describe a mixture of phases, its position would had varied with respect to the free-energy or baryon chemical potential when changing the charge or isospin fraction, as the region with the mixture of phases always encompasses the forced-congruent (no-mixture) coexistence line.

\section*{Acknowledgements}
We acknowledge the colleagues Jeffrey Peterson and Michael Strickland for valuable suggestions. Support for this research comes from the National Science Foundation under grant PHY-1748621, PHAROS (COST Action CA16214), the LOEWE-Program in HIC for FAIR, Conselho Nacional de Desenvolvimento Cient\'{\i}fico e Tecnol\'ogico - CNPq under grant 304758/2017-5 (R.L.S.F), and Funda\c{c}\~ao de Amparo \`a Pesquisa do Estado do Rio Grande do Sul - FAPERGS under grants 19/2551-0000690-0 and 19/2551-0001948-3 (R.L.S.F.).

\appendix
\renewcommand{\theequation}{A\arabic{equation}}
\renewcommand{\thetable}{A\arabic{table}}
\setcounter{equation}{0}  

\section*{Appendix A}

\begin{table}[t!]
\begin{tabular}{ccccc}
\hline
~~~ Particle ~~~    &~~~$Q_{B}$~~~&~~~$Q$~~~&~~~$Q_S$~~~&~~~$Q_I$~~~\\ \hline 
    $p$             &   1         &  1      &   0       &     1/2   \\
    $n$             &   1         &  0      &   0       &    -1/2   \\
    $\Lambda$       &   1         &  0      &   1       &     0     \\
    $\Sigma^+$      &   1         &  1      &   1       &     1     \\
    $\Sigma^0$      &   1         &  0      &   1       &     0     \\
    $\Sigma^-$      &   1         &  -1     &   1       &    -1     \\
    $\Xi^0$         &   1         &  0      &   2       &    -3/2   \\
    $\Xi^-$         &   1         &  -1     &   2       &    -1/2   \\
    $u$             &   1/3       &  2/3    &   0       &     1/2   \\
    $d$             &   1/3       &  -1/3   &   0       &    -1/2   \\
    $s$             &   1/3       &  -1/3   &   1       &     0     \\   \hline
\end{tabular}
\begin{center}
\caption{Baryon number $Q_B$, electric charge $Q$, strangeness $Q_S$, and isospin $Q_I$ quantum numbers for the baryon octet and the three light quarks. Antiparticles carry opposite signs.}
\end{center}
\label{tab:Q}
\end{table}

%***************** FIGURE 5: 3D YI*************************************************************************************
\renewcommand\thefigure{A\arabic{figure}} 
\setcounter{figure}{0}  
\begin{figure*}[t!]
\centering
\begin{subfigure}[b]{0.475\textwidth}
\centering
\includegraphics[width=\textwidth]{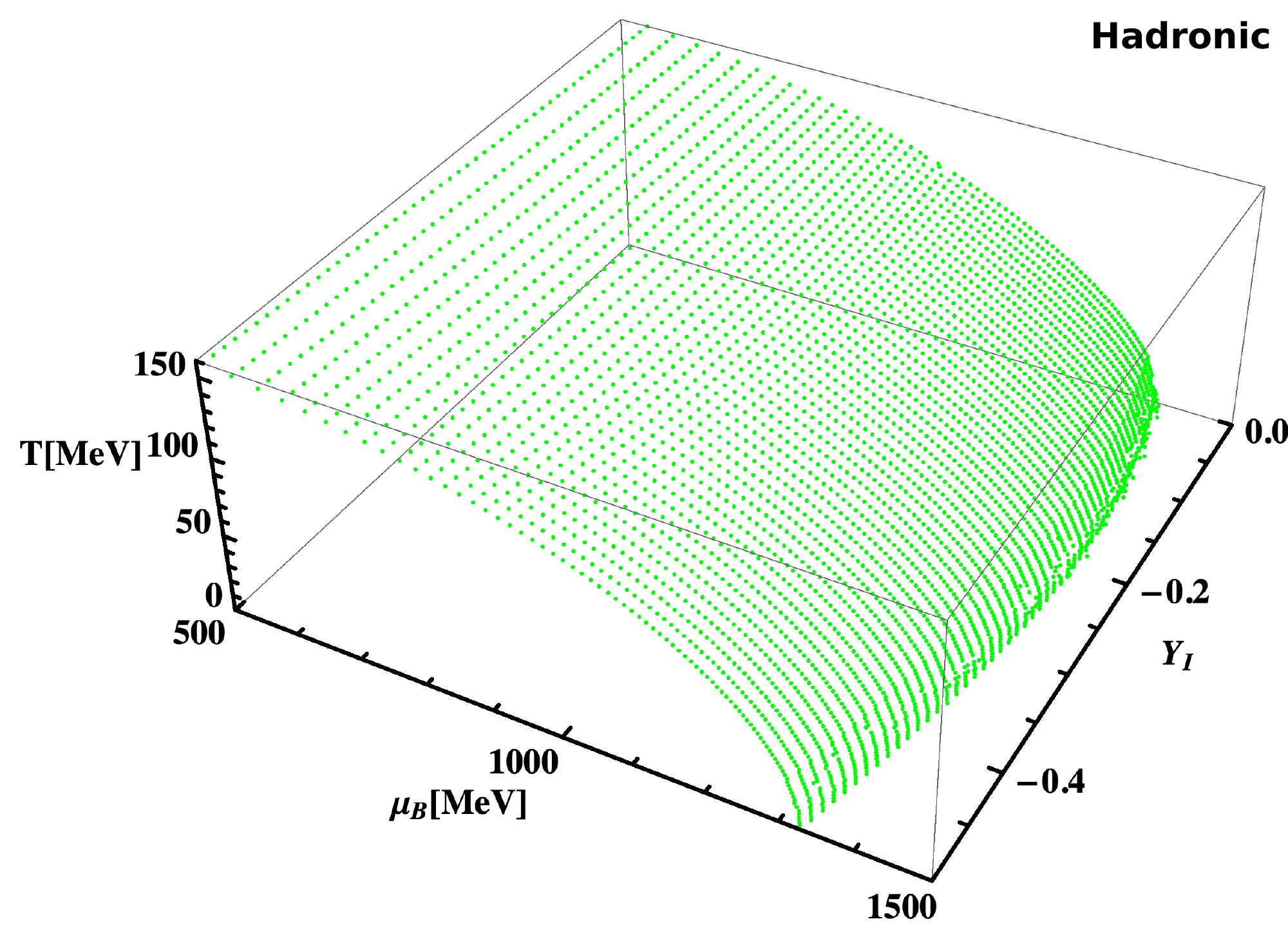}
 \end{subfigure}
 \quad
 \begin{subfigure}[b]{0.475\textwidth} 
\centering
 \includegraphics[width=\textwidth]{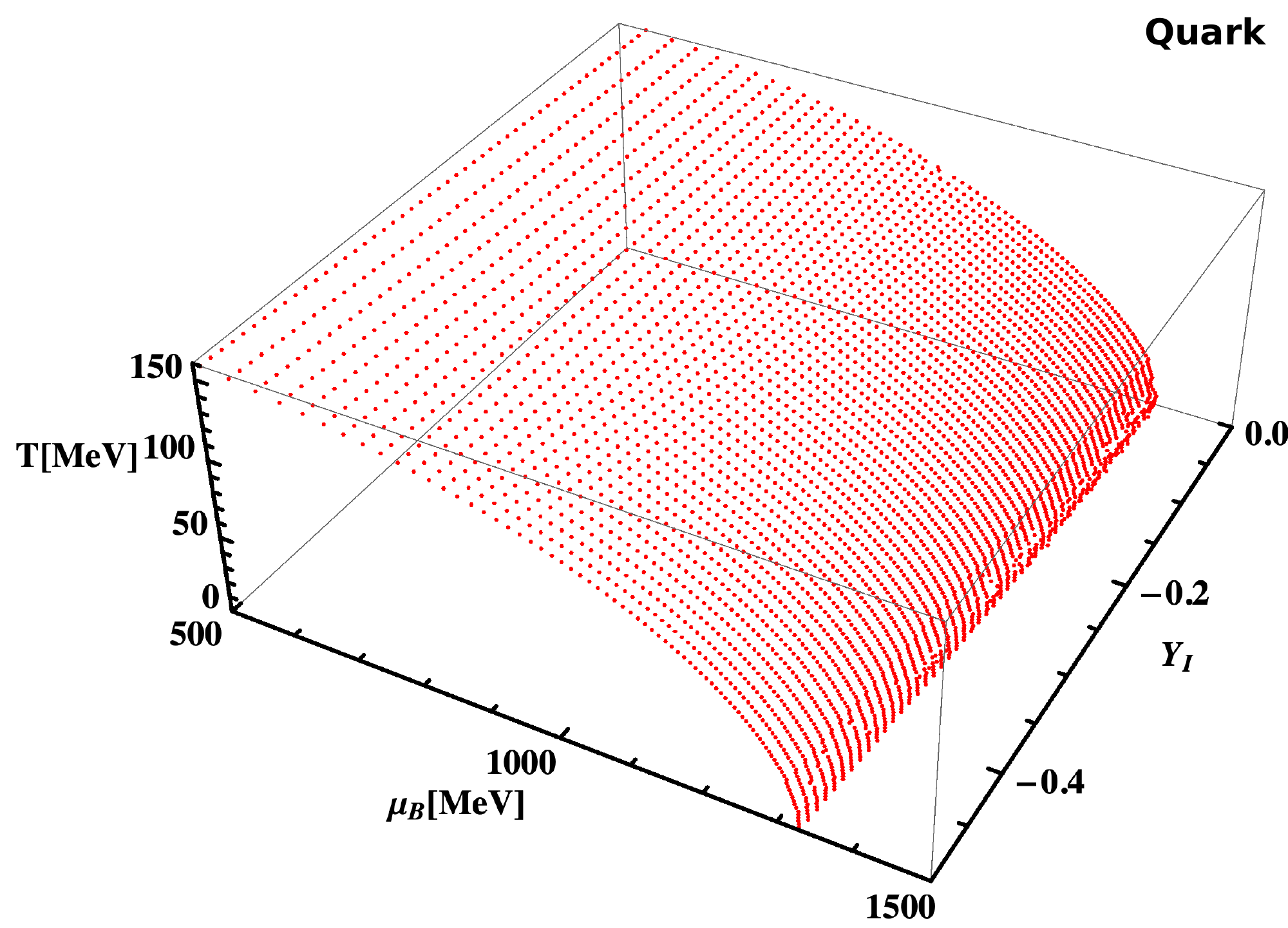}
  \end{subfigure}
\begin{subfigure}[b]{0.475\textwidth}  
\centering
\includegraphics[width=\textwidth]{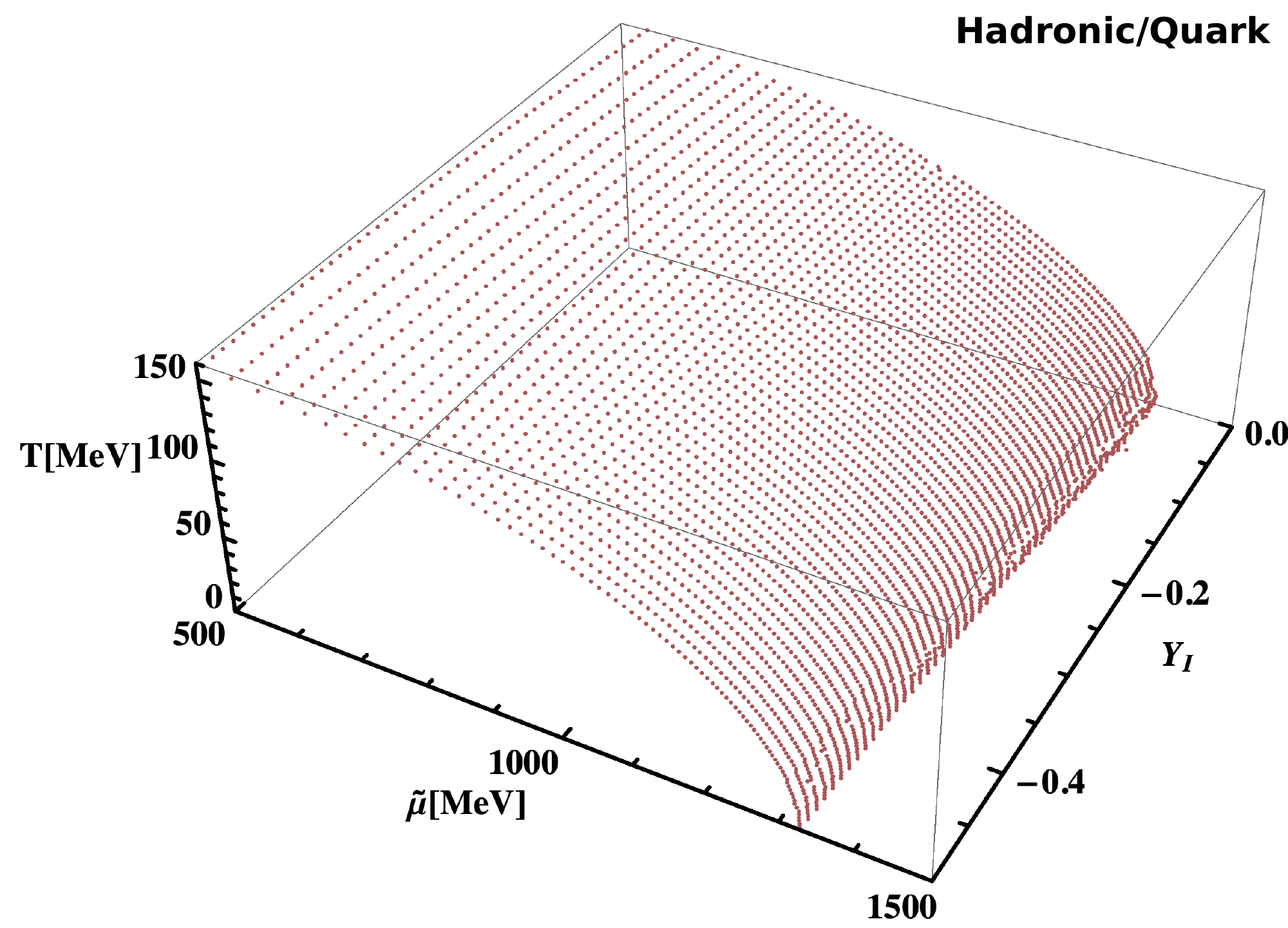}
  \end{subfigure}
\quad
  \caption {Same as Fig.~\ref{fig:fig} but showing the isospin charge fraction $Y_I$.}  
\label{fig3:fig}
\end{figure*}
%******************************************************************************************************************

%***************** FIGURE 6: 3D muI*************************************************************************************
\begin{figure*}[t!]
\centering
\begin{subfigure}[b]{0.475\textwidth}
\centering
\includegraphics[width=\textwidth]{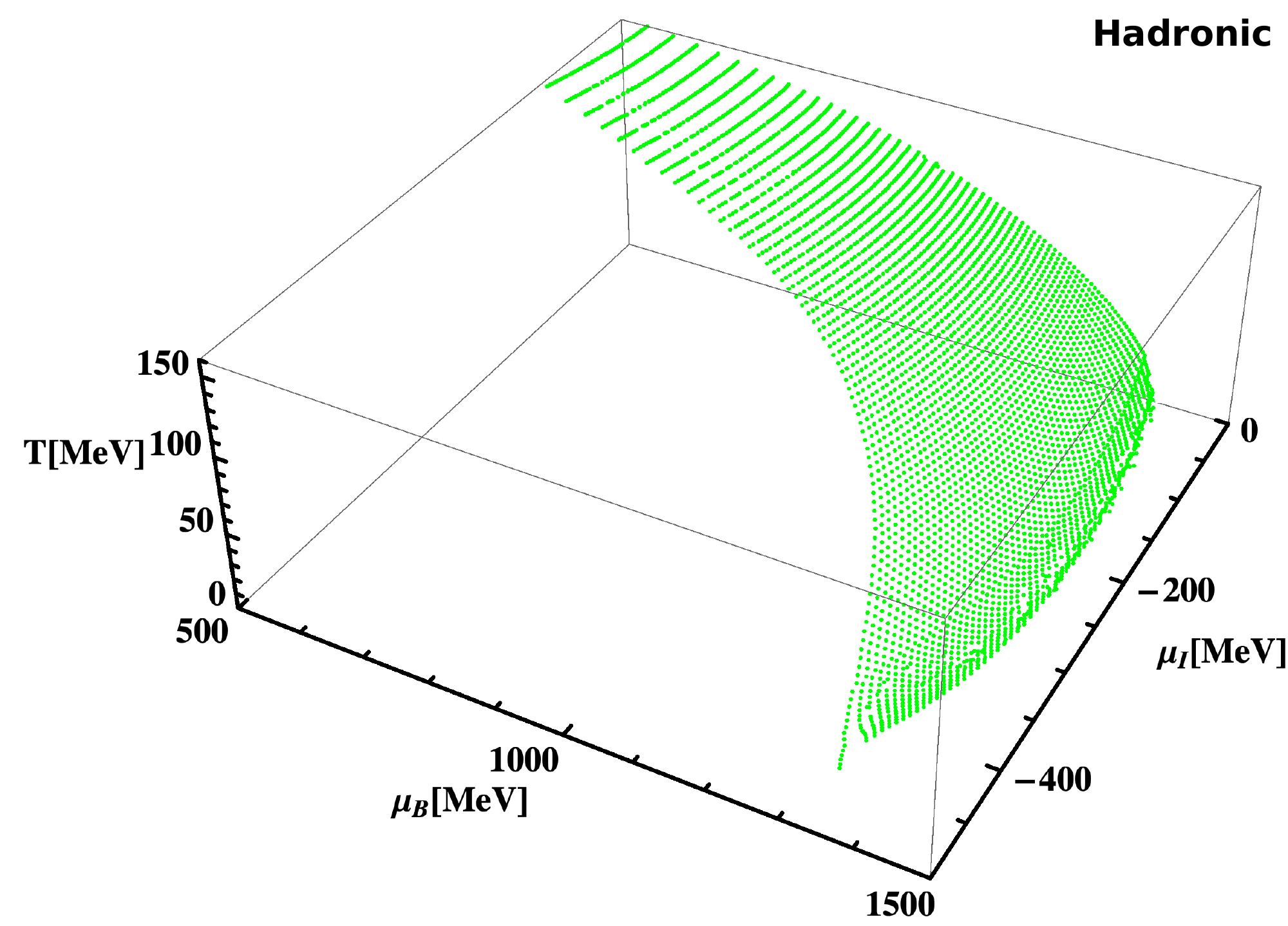}
 \end{subfigure}
 \quad
 \begin{subfigure}[b]{0.475\textwidth} 
\centering
 \includegraphics[width=\textwidth]{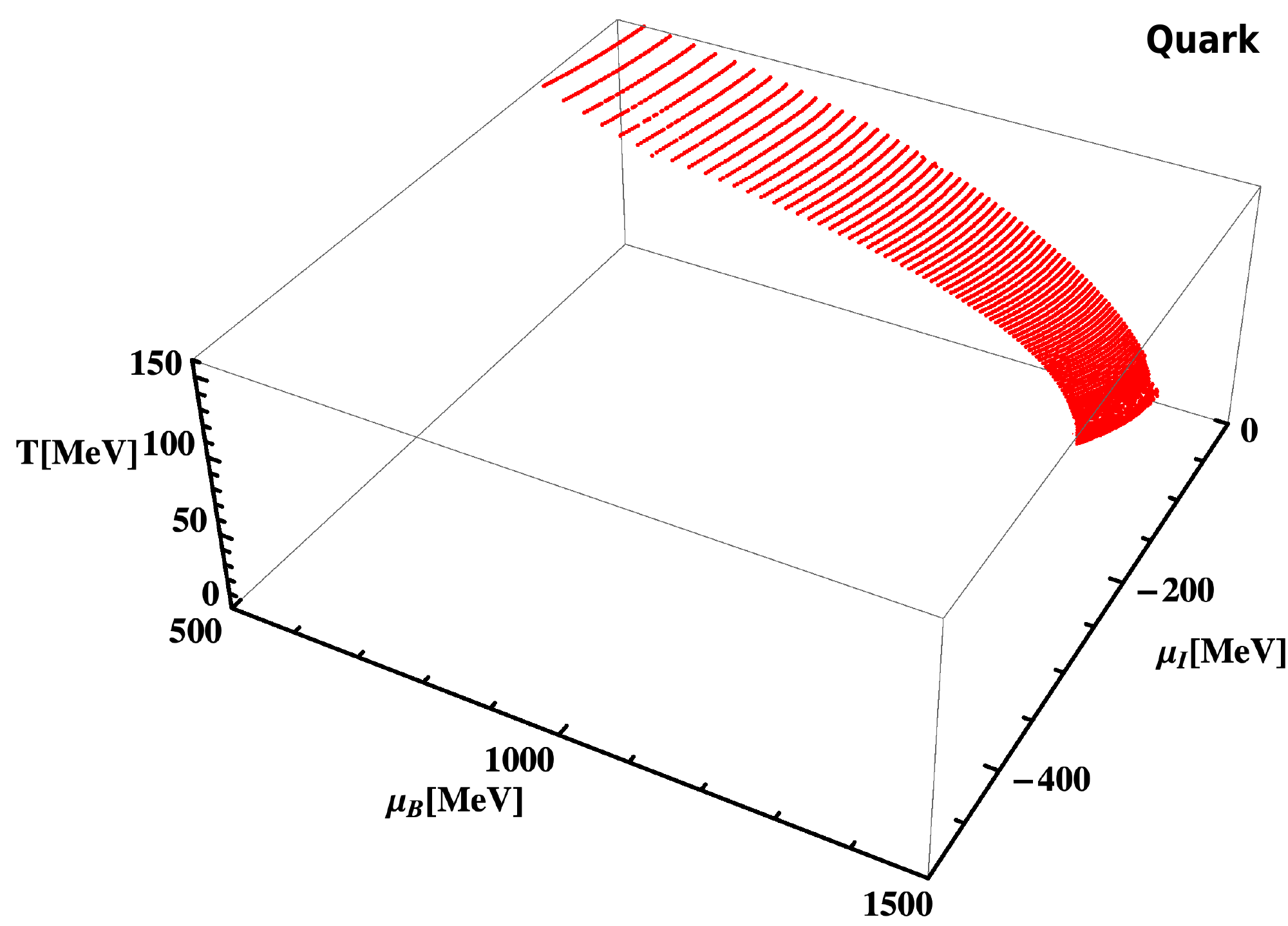}
  \end{subfigure}
\begin{subfigure}[b]{0.475\textwidth}  
\centering
\includegraphics[width=\textwidth]{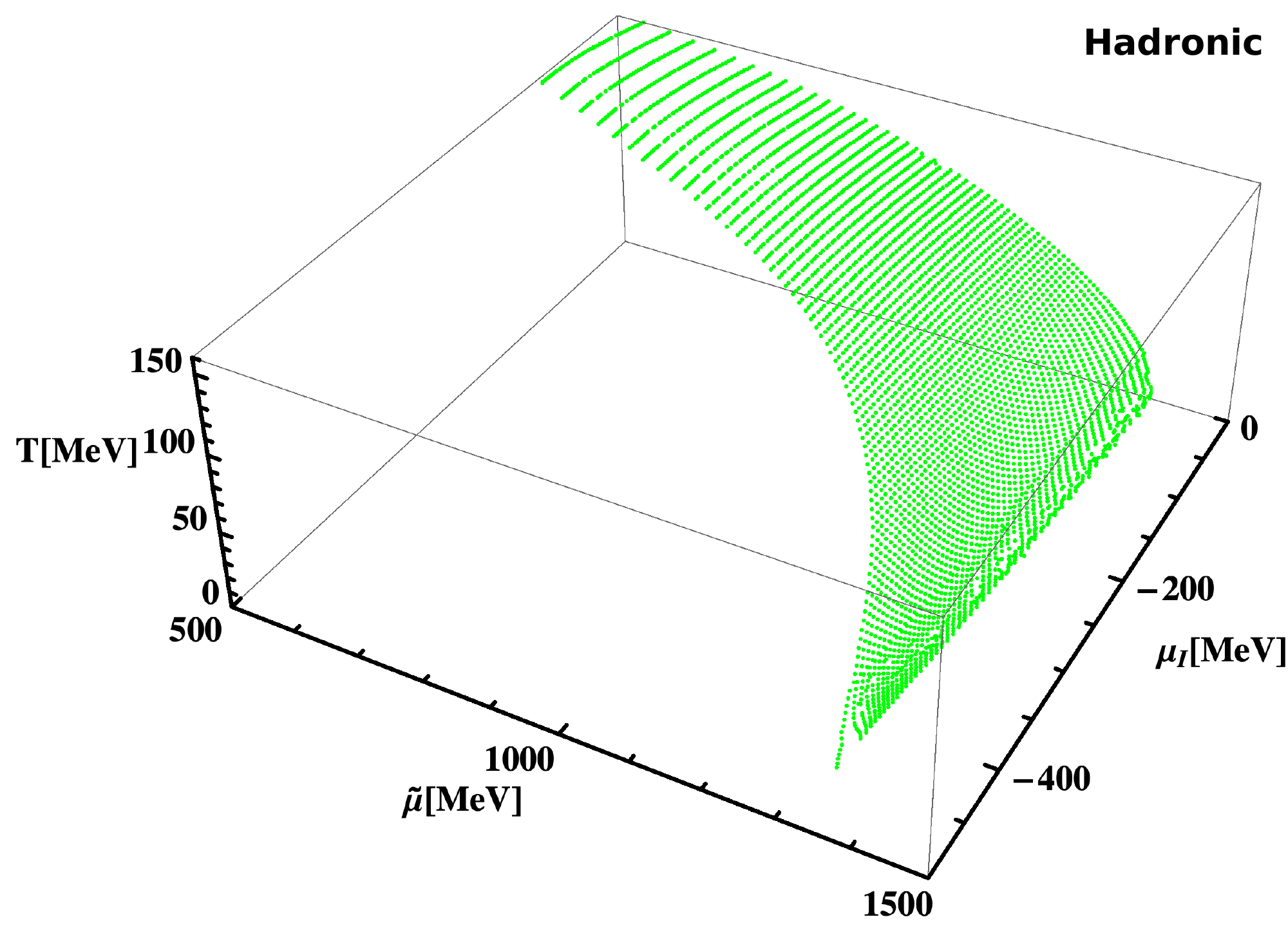}
  \end{subfigure}
\quad
\begin{subfigure}[b]{0.475\textwidth}  
\centering
\includegraphics[width=\textwidth]{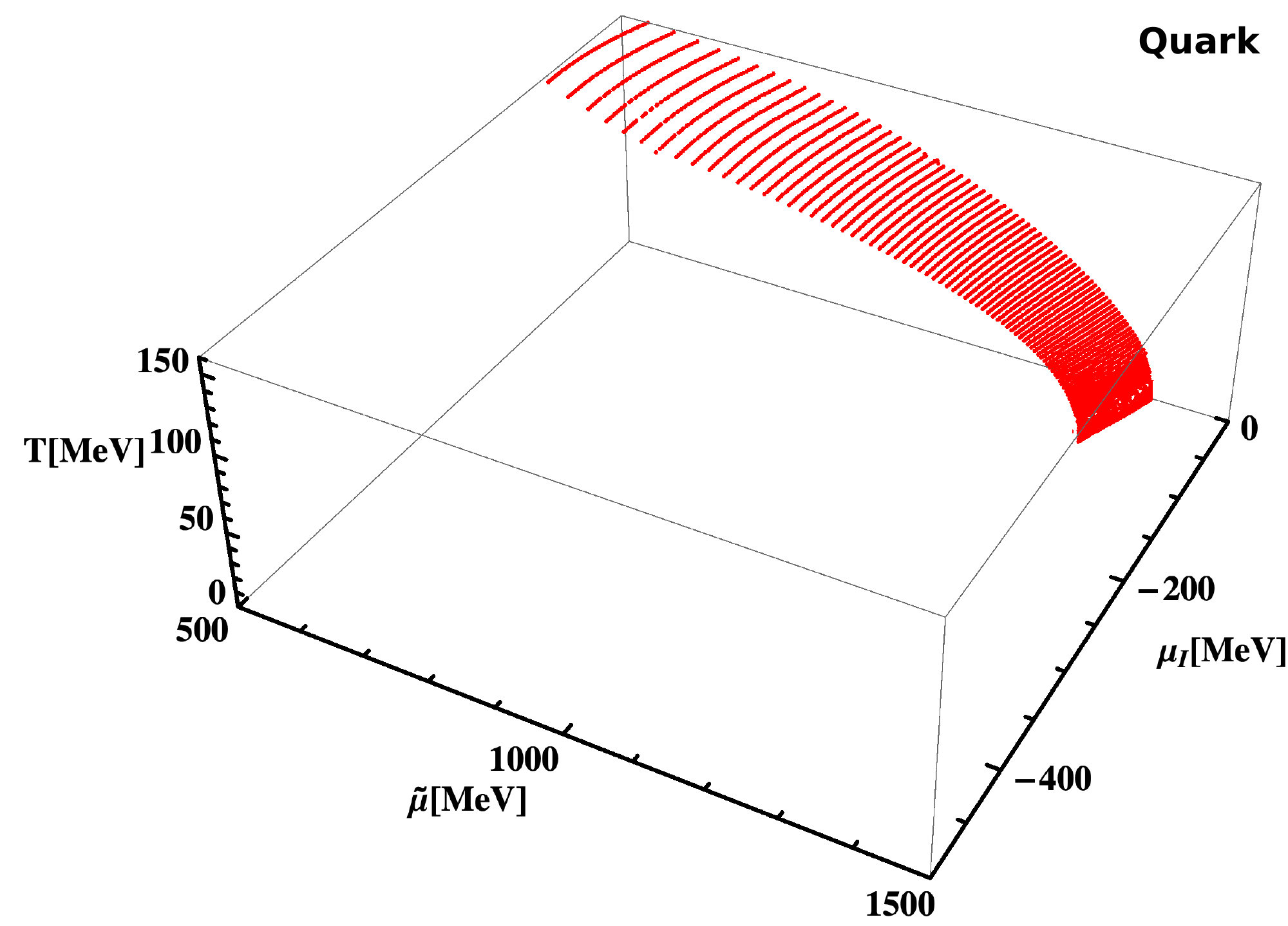}
  \end{subfigure}
  \caption {Same as Fig.~\ref{fig2:fig} but showing the isospin chemical potential $\mu_I$.}    
\label{fig4:fig}
\end{figure*}  
%******************************************************************************************************************

The chemical potentials of the various baryon and quark species are obtained using the appropriate values from Table I in conjunction with Eq. \ref{equation:mui}: 
\begin{eqnarray}
\mu_p &=& \mu_B + \mu_Q ,
\label{equation:mup}
 \nonumber\\
\mu_n &=& \mu_B ,
\label{equation:mun}
\label{equation:mup}
 \nonumber\\
\mu_\Lambda &=& \mu_B + \mu_S ,
\label{equation:mulamda}
\label{equation:mup}
 \nonumber\\
\mu_\Sigma^+ &=& \mu_B + \mu_Q +\mu_S,
\label{equation:musiplus}
\label{equation:mup}
 \nonumber\\
\mu_\Sigma^0 &=& \mu_B + \mu_S ,
\label{equation:musi0}
\label{equation:mup}
 \nonumber\\
\mu_\Sigma^- &=& \mu_B -\mu_Q+ \mu_S ,
\label{equation:musi-}
\label{equation:mup}
 \nonumber\\
\mu_\Xi^0 &=& \mu_B +2\mu_S ,
\label{equation:mup}
 \nonumber\\
\mu_\Xi^- &=& \mu_B - \mu_Q +2\mu_S ,
\label{equation:mup}
\end{eqnarray}
\begin{eqnarray}
\mu_u &=& \frac{1}{3}\mu_B +\frac{2}{3} \mu_Q ,
\label{equation:mup}
 \nonumber\\
\mu_d &=&\frac{1}{3}\mu_B -\frac{1}{3} \mu_Q ,
\label{equation:mup}
 \nonumber\\
\mu_s&=& \frac{1}{3}\mu_B -\frac{1}{3} \mu_Q +\mu_S.
\end{eqnarray}
\

Once more, we remind the reader that we consider the strangeness of particles to be positive in our notation. Otherwise, all $Q_{S,i}$, $n_S$, and $Y_S$ would have to be multiplied by $-1$. This would also reverse the sign of $\mu_S$ in all equations. In the equivalent isospin formalism discussed in Section II B, the chemical potentials for the different species look the same, except for $\mu_Q$ being replaced by $\mu_I$. This can be obtained by replacing the values of $Q_{B,i}$, $Q_{I,i}$ and $Q_{S,i}$ for each baryonic or quark species in Eq.~\eqref{13}. 

\section*{Appendix B}

To extend the discussion of Section III A to the equivalent isospin formalism, we present Figs.~\ref{fig3:fig} and \ref{fig4:fig}, where we plot phase diagrams in terms of the isospin fraction $Y_I$ and isospin chemical potential $\mu_I$ (as opposed to the earlier $Y_Q$ and $\mu_Q$). Since for non-strange matter Eq.~\eqref{vYi} reduces simply to $Y_I=Y_Q-0.5$, Fig.~\ref{fig3:fig} is very similar to Fig.~\ref{fig:fig}, only differing by the $0.5$ shift in the $Y_I$ axis.

Fig.~\ref{fig4:fig} is exactly like Fig.~\ref{fig2:fig}, which is a consequence of the middle term being the same in Eq.~\eqref{ve2} and Eq.~\eqref{lst} in order to reproduce the same particle chemical potential expressions of Appendix A. All of the statements made in this Appendix and at the end of Section III B were verified numerically by rewriting our numerical code to run for fixed isospin fractions. 

\bibliographystyle{apsrev4-1}
\bibliography{paper}% Produces the bibliography via BibTeX.
\end{document}